\documentclass[journal]{IEEEtran} \newcommand{\figsize}{3.0}
\usepackage{graphicx,psfrag}
\usepackage[cmex10]{amsmath}
\usepackage{amsfonts,amssymb,latexsym,hyperref}
\usepackage{cite}
\usepackage[usenames,dvipsnames]{color}

\usepackage{algorithm,algorithmic,setspace}
\usepackage{xpatch} 
\usepackage{xspace} 
\usepackage{etoolbox} 

\newtoggle{pd}
\toggletrue{pd} 

\makeatletter
\def\BState{\State\hskip-\ALG@thistlm}
\xpatchcmd{\algorithmic}{\itemsep\z@}{\itemsep=0.5ex plus2pt}{}{}  
\makeatother

 \newcommand{\putFrag}[4]{\begin{figure}[t]
                            \centering
                            #4
			    \includegraphics[width=#3in,clip]{figures2/#1.eps}
            		    \caption{#2}
           		    \label{fig:#1}
                          \end{figure} }
 
 \newcommand{\capFrag}[2]{}
 \newcommand{\capTable}[2]{}

 \renewcommand{\tilde}{\widetilde}
 \renewcommand{\hat}{\widehat}

 \newcommand{\defn}{\triangleq}
 
 \newcommand{\uvec}[1]{\ensuremath{\underline{\boldsymbol{#1}}}}
 \newcommand{\tvec}[1]{\ensuremath{\Tilde{\boldsymbol{#1}}}}
 \newcommand{\ovec}[1]{\ensuremath{\Bar{\boldsymbol{#1}}}}
 \newcommand{\hvec}[1]{\ensuremath{\Hat{\boldsymbol{#1}}}}
    
 \renewcommand{\vec}[1]{\ensuremath{\boldsymbol{#1}}}
 \newcommand{\mat}[1]{\ensuremath{\begin{bmatrix}#1\end{bmatrix}}}

 \newcommand{\mc}[1]{\ensuremath{\mathcal{#1}}}

 \newcommand{\Real}{{\mathbb{R}}}
 \newcommand{\Complex}{{\mathbb{C}}}
 \newcommand{\Int}{{\mathbb{Z}}}

 \newcommand{\of}[1]{^{(#1)}}
 
 \newcommand{\ofc}[1]{^{{(#1)}*}}
 \newcommand{\ofH}[1]{^{{(#1)}\textsf{H}}}
 \newcommand{\tran}{^{\text{\textsf{T}}}}
 \newcommand{\herm}{^{\text{\textsf{H}}}}
 
 \newcommand*\dif{\mathop{}\!\mathrm{d}} 

 \DeclareMathOperator{\E}{\mathbb{E}}

 \DeclareMathOperator{\Cov}{Cov}

 \DeclareMathOperator{\rank}{rank}
 \DeclareMathOperator{\tr}{tr}
 \DeclareMathOperator{\diag}{diag}

 \DeclareMathOperator{\vect}{vec}
 
 \DeclareMathOperator{\Diag}{Diag}
 
 \DeclareMathOperator*{\argmin}{arg\,min}
 \DeclareMathOperator*{\argmax}{arg\,max}

 \renewcommand{\eqref}[1]{(\ref{eq:#1})}
 
 \newcommand{\Figref}[1]{Figure~\ref{fig:#1}}
 \newcommand{\figref}[1]{Fig.~\ref{fig:#1}}
 
 \newcommand{\secref}[1]{Sec.~\ref{sec:#1}}

 \newcommand{\algref}[1]{Alg.~\ref{alg:#1}}
 \newcommand{\Algref}[1]{Algorithm~\ref{alg:#1}}
 \newcommand{\lineref}[1]{line~\ref{line:#1}}

 \newcommand{\textb}[1]{\textcolor{black}{#1}}
 \newcommand{\blue}{\color{black}}

 \newcounter{comment}[section]
 
 \newcounter{texthead}[section]
 
\newcommand{\ml}{_{\text{\sf ML}}}

\newcommand{\Sigset}{\mc{S}}
\newcommand{\train}{_{\text{\textsf{t}}}}
\newcommand{\trainone}{_{\text{\textsf{t}},1}}
\newcommand{\trainm}{_{\text{\textsf{t}},m}}
\newcommand{\trainmprime}{_{\text{\textsf{t}},m'}}
\newcommand{\trainM}{_{\text{\textsf{t}},M}}
\newcommand{\data}{_{\text{\textsf{d}}}}
\newcommand{\pulse}{_{\text{\textsf{p}}}}

\newcommand{\notl}{_{-l}}

\newcommand{\pfa}{$10^{-3}$}
\iftoggle{pd}{
\newcommand{\metric}{detection-rate at false-alarm-rate=\pfa\xspace}
\newcommand{\Metric}{Detection-rate at false-alarm-rate=\pfa\xspace}
}{
\newcommand{\metric}{detection error rate\xspace}
\newcommand{\Metric}{Detection error rate\xspace}
}

\begin{document}
\setlength{\arraycolsep}{0.5mm}
 \title{Adaptive Detection of Structured Signals in Low-Rank Interference}
 \author{Philip Schniter,\IEEEauthorrefmark{1} \IEEEmembership{Fellow,~IEEE},
         and Evan Byrne%
 \thanks{P.~Schniter and E.~Byrne 
         (email: schniter.1@osu.edu and byrne.133@osu.edu) are with
         the Department of Electrical and Computer Engineering,
         The Ohio State University, Columbus, OH, 43210, USA\@.
         Please direct all correspondence to Prof. Philip Schniter, 
         Dept. ECE, 2015 Neil Ave., Columbus, OH 43210, USA, phone 614.247.6488, fax 614.292.7596.}
 \thanks{This work was supported by NSF grant 1716388 and MIT Lincoln Labs.}
 }
 \date{\today}
 \maketitle

\begin{abstract}
In this paper, we consider the problem of detecting the presence (or absence) of an unknown but structured signal from the space-time outputs of an array under strong, non-white interference.  
Our motivation is the detection of a communication signal in jamming, where often the training portion is known but the data portion is not. 
We assume that the measurements are corrupted by additive white Gaussian noise of unknown variance and a few strong interferers, whose number, powers, and array responses are unknown.
We also assume the desired signal's array response is unknown.
To address the detection problem, we propose several GLRT-based detection schemes that employ a probabilistic signal model and use the EM algorithm for likelihood maximization.
Numerical experiments are presented to assess the performance of the proposed schemes.
\end{abstract}


\section{Introduction}              

\subsection{Problem statement} \label{sec:problem}

Consider the problem of detecting the presence or absence of a signal $\vec{s}\in\Complex^{L}$ from the measured output $\vec{Y}\in\Complex^{M\times L}$ of an $M$-element antenna array.
We are interested in the case where $\vec{s}$ is unknown but structured.
\textb{A motivating example arises with communications signals, where typically a few ``training'' symbols are known and the remaining ``data'' symbols are unknown, apart from their alphabet.}
We will assume that the signal's array response $\vec{h}\in\Complex^M$ is completely unknown but constant over the measurement epoch and signal bandwidth.
The complete lack of knowledge about $\vec{h}$ is appropriate when the array manifold is unknown or uncalibrated (e.g., see the discussion in \cite{Forsythe:LLJ:97}), or when the signal is observed in a dense multipath environment (e.g., \cite{Bliss:TSP:10}).
Also, we will assume that the measurements are corrupted by white noise of unknown variance and $N\geq 0$ possibly strong interferers.
The interference statistics are assumed to be unknown, as is $N$.

The signal-detection problem can be formulated as a binary hypothesis test \cite{Scharf:Book:91} between hypotheses $\mc{H}_1$ (signal present) and $\mc{H}_0$ (signal absent), i.e.,
\begin{subequations}
\label{eq:H1H0} 
\begin{align}
\mc{H}_1: \vec{Y}
&= \vec{h}\vec{s}\herm + \vec{B\Phi}\herm + \vec{W} \in \Complex^{M\times L} 
\label{eq:H1}\\
\mc{H}_0: \vec{Y}
&= \vec{B\Phi}\herm + \vec{W} \in \Complex^{M\times L} 
\label{eq:H0} .
\end{align}
\end{subequations}
In \eqref{H1H0}, 
$\vec{W}$ refers to the noise and
$\vec{B\Phi}\herm$ to the interference.
We model $\vec{W}$ as white Gaussian noise \textb{(WGN)}\footnote{%
\textb{By white Gaussian, we mean that $\vec{W}$ has i.i.d.\ zero-mean circularly symmetric complex Gaussian entries.}
} with unknown variance $\nu>0$.
If the array responses of the $N$ interferers are constant over the measurement epoch and bandwidth, then the rank of $\vec{B\Phi}\herm$ will be at most $N$.
As will be discussed in the sequel, we will sometimes (but not always) model the temporal interference component $\vec{\Phi}\herm$ as white and Gaussian.

Communications signals often take a form like
\begin{align}
\vec{s}\herm
&= \mat{\vec{s}\train\herm & \vec{s}\data\herm}
\label{eq:std},
\end{align}
where $\vec{s}\train\in\Complex^Q$ is a known training sequence,
$\vec{s}\data\in\mc{A}^{L-Q}$ is an unknown data sequence, $\mc{A}\subset\Complex$ is a finite alphabet, and $Q\ll L$.
Suppose that the measurements are partitioned as $\vec{Y}=\mat{\vec{Y}\train&\vec{Y}\data}$, conformal with \eqref{std}.
For the purpose of signal detection or synchronization, the data measurements $\vec{Y}\data$ are often ignored (see, e.g., \cite{Bliss:TSP:10}).
But these data measurements can be very useful, especially when the training symbols (and thus the training measurements $\vec{Y}\train$) are few.
Our goal is to develop detection schemes that use all measurements $\vec{Y}$ while handling the incomplete knowledge of $\vec{s}$ in a principled manner.

We propose to model the signal structure probabilistically.
That is, we treat $\vec{s}$ as a random vector with prior pdf $p(\vec{s})$,
\textb{where $\vec{s}$ is statistically independent of $\vec{h}$, $\vec{B}$, $\vec{\Phi}$, and $\vec{W}$.} 
Although the general methodology we propose supports arbitrary $p(\vec{s})$, we sometimes focus (for simplicity) on the case of statistically independent components, i.e., 
\begin{align}
p(\vec{s})
&= \prod_{l=1}^L p_l(s_l)
\label{eq:s_indep} .
\end{align}

For example, with uncoded communication signals partitioned as in \eqref{std}, we would use \eqref{s_indep} with
\begin{align}
p_l(s_l)
&= \begin{cases}
\delta(s_l-s_{\textsf{t},l}) & l=1,\dots,Q\\
\frac{1}{|\mc{A}_l|}\sum_{s\in\mc{A}_l} \delta(s_l-s) & l=Q+1,\dots,L,
\end{cases}
\label{eq:std_indep}
\end{align}
where $\delta(\cdot)$ denotes the Dirac delta, 
\textb{$s_{\textsf{t},l}$ the $l$th training symbol,
$\mc{A}_l$ is a finite-cardinality set containing the $l$th data symbol,
and $|\mc{A}_l|$ is the cardinality of $\mc{A}_l$.}
For coded communications signals, the independent prior \eqref{s_indep} would still be appropriate if a ``turbo equalization'' \cite{Koetter:SPM:04} approach was used, where symbol estimation is iterated with soft-input soft-input decoding.
A variation of \eqref{std} that avoids the need to know $\mc{A}$ follows from modeling $\{s_l\}_{l=Q+1}^L$ as i.i.d.\ Gaussian.
\textb{In practical communications scenarios, there exists imperfect time and frequency synchronization, which leads to mismatch in the assumed model \eqref{s_indep}-\eqref{std_indep}.
In \secref{num}, we discuss synchronization mismatch and investigate its effect in numerical experiments.}

The proposed probabilistic framework is quite general.
For example, in addition to training/data structures of the form in \eqref{std}, the independent model \eqref{s_indep} covers superimposed training \cite{Tong:SPM:04}, bit-level training \cite{Schniter:JSTSP:11}, constant-envelope waveforms \cite{Forsythe:LLJ:97}, and pulsed signals (i.e., $\vec{s}\herm=\mat{\vec{s}\pulse\herm & \vec{0}\herm}$ with unknown $\vec{s}\pulse$) \cite{Forsythe:LLJ:97}.
To exploit sinusoidal signal models, or signals with known spectral characteristics (see, e.g., \cite{Forsythe:LLJ:97}), the independent model \eqref{s_indep} would be discarded in favor of a more appropriate $p(\vec{s})$.
\textb{There is an excellent description of most of these topics in \cite{Forsythe:LLJ:97}, and we refer readers to that source for more details.}


\subsection{Prior work} \label{sec:prior}

For the case where the entire signal $\vec{s}\in\Complex^L$ is \emph{known}, the detection problem \eqref{H1H0} has been studied in detail.
For example, in the classical work of Kelly \cite{Kelly:TAES:86,Kelly:Tech:89}, the interference-plus-noise $\vec{B\Phi}\herm + \vec{W}$ was modeled as temporally white\footnote{\label{white}%
\textb{By temporally white and Gaussian, we mean that the columns are i.i.d.\ circularly symmetric complex Gaussian random vectors with zero mean and a generic covariance matrix.}}
and Gaussian with unknown (and unstructured) spatial covariance $\vec{\Sigma}>0$, and the generalized likelihood ratio test (GLRT) \cite{Scharf:Book:91} was derived.
Detector performance can be improved when the interference is known to have low rank. 
For example, Gerlach and Steiner \cite{Gerlach:TSP:00} assumed temporally white Gaussian interference with known noise variance $\nu$ and unknown interference rank $N$ and derived the GLRT.
More recently, Kang, Monga, and Rangaswamy \cite{Kang:TAES:14} assumed temporally white Gaussian interference with unknown $\nu$ and known $N$ and derived the GLRT.
Other structures on $\vec{\Sigma}$ were considered by Aubry et al.\ in \cite{Aubry:TSP:13}.
In a departure from the above methods, McWhorter \cite{McWhorter:ASAP:04} proposed to treat the interference components $\vec{B}\in\Complex^{M\times N}$ and $\vec{\Phi}\in\Complex^{L\times N}$, as well as the noise variance $\nu$, as deterministic unknowns. 
He then derived the corresponding GLRT.
Note that McWhorter's approach implicitly assumes knowledge of the interference rank $N$.
Bandiera et al.~\cite{Bandiera:TSP:07} proposed yet a different approach, based on a Bayesian perspective.

For adaptive detection of \emph{unknown} but structured signals $\vec{s}$, we are aware of relatively little prior work.
Forsythe \cite[p.110]{Forsythe:LLJ:97} describes an iterative scheme for signals with deterministic (e.g., finite-alphabet, constant envelope) structure that builds on Kelly's GLRT.
Each iteration involves maximum-likelihood (ML) signal estimation and least-squares beamforming, based on the intuition that correct decisions will lead to better beamformers and thus better interference suppression.
Error propagation remains a serious issue, however, as we will demonstrate in the sequel.

\subsection{Contributions}

We propose three GLRT-based schemes for adaptive detection of unknown structured signals $\vec{s}$ with unknown array responses $\vec{h}$, \textb{additive WGN} of unknown variance $\nu$, and interference $\vec{B\Phi}\herm$ of possibly low rank.
All of our schemes use a probabilistic signal model $\vec{s}\sim p(\vec{s})$, under which the direct evaluation of the GLRT numerator becomes intractable.
To circumvent this intractability, we use expectation maximization (EM) \cite{Dempster:JRSS:77}.
In particular, we derive computationally efficient EM procedures for the independent prior \eqref{s_indep}, paying special attention to finite-alphabet and Gaussian cases.

Our first approach treats the interference $\vec{B\Phi}\herm$ as \textb{temporally white$^{\footnotesize\ref{white}}$ and Gaussian}, and it makes no attempt to leverage low interference rank, similar to Kelly's approach \cite{Kelly:TAES:86}.
A full-rank interference model would be appropriate if, say, the interferers' array responses varied significantly over the measurement epoch.
We show that our first approach is a variation on Forsythe's iterative scheme \cite[p.110]{Forsythe:LLJ:97} that uses ``soft'' symbol estimation and ``soft'' signal subtraction, making it much more robust to error propagation.

Our second approach is an extension of our first that aims to exploit the possibly low-rank nature of the interference.
As in \cite{Gerlach:TSP:00,Kang:TAES:14,Aubry:TSP:13},
the interference is modeled as temporally white Gaussian but, different from \cite{Gerlach:TSP:00,Kang:TAES:14,Aubry:TSP:13}, both the interference rank $N$ and the noise variance $\nu$ are unknown.
More significantly, unlike \cite{Gerlach:TSP:00,Kang:TAES:14,Aubry:TSP:13}, the signal $\vec{s}$ is assumed to be unknown.

Our third approach also aims to exploit low-rank interference, but it does so while modeling the interference as deterministic, as in McWhorter \cite{McWhorter:ASAP:04}.
Unlike \cite{McWhorter:ASAP:04}, however, the interference rank $N$ and the signal $\vec{s}$ are assumed to be unknown.
Numerical experiments are presented to demonstrate the efficacy of our three approaches.

\section{Background} \label{sec:back}

We first provide some background that will be used in developing the proposed methods.
In our discussions below, we will use $\vec{P}_{\vec{A}}$ to denote orthogonal projection onto the column space of a given matrix $\vec{A}$, i.e.,
\begin{align}
\vec{P}_{\vec{A}} 
\defn \vec{A}(\vec{A}\herm\vec{A})^{-1}\vec{A}\herm ,
\end{align}
and $\vec{P}_{\vec{A}}^\perp\defn \vec{I}-\vec{P}_{\vec{A}}$ to denote the orthogonal complement.
Recall that both $\vec{P}_{\vec{A}}$ and $\vec{P}_{\vec{A}}^\perp$ are Hermitian and idempotent. 

\subsection{Full-rank Gaussian Interference}

The classical work of Kelly \cite{Kelly:TAES:86,Kelly:Tech:89} tackled the binary hypothesis test \eqref{H1H0} by treating the interference-plus-noise $\textb{\vec{N}\defn} \vec{B\Phi}\herm + \vec{W}$ as temporally white and Gaussian with unknown $\textb{M\times M}$ spatial covariance matrix $\vec{\Sigma}>0$. 
This reduces \eqref{H1H0} to
\begin{subequations}
\label{eq:H1H0_Sigma} 
\begin{align}
\mc{H}_1: \vec{Y}
&= \vec{h}\vec{s}\herm + 
\textb{
\vec{N} \text{~for~}
\vect(\vec{N})\sim\mc{CN}(\vec{0},\vec{I}_L\otimes\vec{\Sigma})
}\\
\mc{H}_0: \vec{Y}
&= 
\textb{
\vec{N} \text{~for~}
\vect(\vec{N})\sim\mc{CN}(\vec{0},\vec{I}_L\otimes\vec{\Sigma})
} ,
\end{align}
\end{subequations}
\textb{where 
$\vect(\vec{N})$ denotes the vector formed by concatenating all columns of the matrix $\vec{N}$, 
$\mc{CN}(\vec{\mu},\vec{C})$ denotes the circularly symmetric multivariate complex Gaussian distribution with mean vector $\vec{\mu}$ and covariance matrix $\vec{C}$,
and $\otimes$ denotes the Kronecker product.
We note that the covariance structure $\vec{I}_L\otimes\vec{\Sigma}$ in \eqref{H1H0_Sigma} corresponds temporal whiteness across $L$ time samples and spatial correlation with covariance matrix $\vec{\Sigma}$.}
With known $\vec{s}$, the GLRT \cite{Scharf:Book:91} takes the form
\begin{align}
\frac{\max_{\vec{h},\vec{\Sigma}>0} p(\vec{Y}|\mc{H}_1;\vec{h},\vec{\Sigma})}
{\max_{\vec{\Sigma}>0} p(\vec{Y}|\mc{H}_0;\vec{\Sigma})} 
\gtreqless \eta
\label{eq:glrt_kelly},
\end{align}
for some threshold $\eta$. 
Using results from \cite{Goodman:AMS:63}, it was shown in \cite{Kelly:TAES:86} that \eqref{glrt_kelly} reduces to
\begin{align}
\frac{\prod_{m=1}^M\lambda_{0,m}}{\prod_{m=1}^M\lambda_{1,m}}
\gtreqless \eta 
\label{eq:glrt_kelly2},
\end{align}
for decreasing ordered (i.e., $\lambda_{i,m}\geq \lambda_{i,m+1}~\forall m,i$) eigenvalues
\begin{subequations}
\begin{align}
\{\lambda_{0,m}\}_{m=1}^M &\defn \text{eigenvalues}\big(\tfrac{1}{L}\vec{YY}\herm\big)\\ 
\{\lambda_{1,m}\}_{m=1}^M &\defn \text{eigenvalues}\big(\tfrac{1}{L}\vec{YP}_{\vec{s}}^\perp\vec{Y}\herm\big) .
\end{align}\label{eq:lambda}%
\end{subequations}
Kelly's approach was applied to the detection/synchronization of communications signals by Bliss and Parker in \cite{Bliss:TSP:10} after discarding the measurements corresponding to the unknown data symbols $\vec{s}\data$.

When $L<M+1$, some eigenvalues will be zero-valued and so the test \eqref{glrt_kelly2} is not directly applicable.
One can imagine many strategies to circumvent this problem (e.g., restricting to positive eigenvalues, computing eigenvalues from a regularized sample covariance of the form $\frac{b}{L}\vec{YY}\herm + c\vec{I}$ for $b,c>0$, etc) that can be considered as departures from Kelly's approach.
In the sequel, we describe approaches that use a low-rank-plus-identity covariance $\vec{\Sigma}$, as would be appropriate when the interferers are few, i.e., $N\ll M$.

\subsection{Low-rank Gaussian Interference}

The low-rank property of the interference $\vec{B\Phi}\herm$ can be exploited to improve detector performance.
Some of the first work in this direction was published by Gerlach and Steiner in \cite{Gerlach:TSP:00}.
They assumed known noise variance $\nu$ and temporally white Gaussian interference, so that \textb{$\vect(\vec{B\Phi}\herm+\vec{W})\sim\mc{CN}(\vec{0},\vec{I}_L\otimes\vec{\Sigma})$ where $\vec{\Sigma}=\vec{R}+\nu\vec{I}_M$} with unknown low-rank $\vec{R}\geq 0$.
The GLRT was then posed under the constraint that $\vec{\Sigma}\in \Sigset_\nu \defn \{\vec{R}+\nu\vec{I}: \vec{R}\geq 0\}$:
\begin{align}
\frac{\max_{\vec{h},\vec{\Sigma}\in\Sigset_\nu} p(\vec{Y}|\mc{H}_1;\vec{h},\vec{\Sigma})}
{\max_{\vec{\Sigma}\in\Sigset_\nu} p(\vec{Y}|\mc{H}_0;\vec{\Sigma})} 
\gtreqless \eta
\label{eq:glrt_gs}.
\end{align}
They showed that the GLRT \eqref{glrt_gs} reduces to one of the form \eqref{glrt_kelly2}, but with thresholded eigenvalues $\tilde{\lambda}_{i,m}=\max\{\lambda_{i,m},\nu\}$.


More recently, Kang, Monga, and Rangaswamy \cite{Kang:TAES:14} proposed a variation on Gerlach and Steiner's approach \cite{Gerlach:TSP:00} where the noise variance $\nu$ is unknown but $N=\rank(\vec{R})$ is known, $N<M$, \textb{and $N\leq L$.}
In particular, they proposed the GLRT 
\begin{align}
\frac{\max_{\vec{h},\vec{\Sigma}\in\Sigset_N} p(\vec{Y}|\mc{H}_1;\vec{h},\vec{\Sigma})}
{\max_{\vec{\Sigma}\in\Sigset_N} p(\vec{Y}|\mc{H}_0;\vec{\Sigma})} 
\gtreqless \eta
\label{eq:glrt_kmr} ,
\end{align}
where 
\begin{align}
\Sigset_N 
&\defn \{\vec{R}+\nu\vec{I}: \rank(\vec{R})=N, \vec{R}\geq 0, \nu>0\}
\label{eq:SigsetN} .
\end{align}
Using a classical result from \cite{Anderson:AMS:63}, it can be shown that the GLRT \eqref{glrt_kmr} simplifies to 
\begin{align}
\frac{\prod_{m=1}^M \hat{\lambda}_{0,m}}{\prod_{m=1}^M\hat{\lambda}_{1,m}}
\gtreqless \eta 
\label{eq:glrt_kmr2} ,
\end{align}
with $\{\hat{\lambda}_{i,m}\}_{m=1}^{M}$ a smoothed version of $\{\lambda_{i,m}\}_{m=1}^M$ from \eqref{lambda}:
\begin{align}
\hat{\lambda}_{i,m} &\defn 
\begin{cases}
\lambda_{i,m} & m=1,\dots,N, \\
\hat{\nu}_i & m=N+1,\dots,M. \\
\end{cases}
\label{eq:lamihata}\\
\hat{\nu}_i &\defn \frac{1}{M-N}\sum_{m=N+1}^M \lambda_{i,m} 
\label{eq:vihata} .
\end{align}


\subsection{Low-rank Deterministic Interference}

The approaches discussed above all model the interference $\vec{B\Phi}\herm$ as temporally white Gaussian. 
McWhorter~\cite{McWhorter:ASAP:04} instead proposed to treat the interference components $\vec{B}\in\Complex^{M\times N}$ and $\vec{\Phi}\in\Complex^{L\times N}$ as deterministic unknowns, yielding the
GLRT
\begin{align}
\frac{\max_{\vec{h},\vec{B},\vec{\Phi},\nu>0} p(\vec{Y}|\mc{H}_1;\vec{h},\vec{B},\vec{\Phi},\nu)}
{\max_{\vec{B},\vec{\Phi},\nu>0} p(\vec{Y}|\mc{H}_0;\vec{B},\vec{\Phi},\nu)} 
\gtreqless \eta
\label{eq:glrt_mc},
\end{align}
where the interference rank $N$ is implicitly known.
It was shown in \cite{McWhorter:ASAP:04} that the GLRT \eqref{glrt_mc} simplifies to 
\begin{align}
\frac{\hat{\nu}_0}{\hat{\nu}_1}
= \frac{\sum_{m=N+1}^M\lambda_{0,m}}{\sum_{m=N+1}^M\lambda_{1,m}}
\gtreqless \eta' 
\label{eq:glrt_mc2}
\end{align}
using the $\{\lambda_{i,m}\}$ defined in \eqref{lambda}.
Comparing \eqref{glrt_mc2} to \eqref{glrt_kmr2}, we see that both GLRTs involve noise variance estimates $\hat{\nu}_i$ computed by averaging the smallest eigenvalues.
However, \eqref{glrt_mc2} discards the largest $N$ eigenvalues whereas \eqref{glrt_kmr2} uses them in the test.

\section{GLRTs via White Gaussian Interference} \label{sec:GerlachSteiner}

We now consider adaptive detection via the binary hypothesis test \eqref{H1H0} with unknown structured $\vec{s}\in\Complex^L$. 
As described earlier, our approach is to model $\vec{s}$ as a random vector with prior density $p(\vec{s})$.

Our first approach treats the interference $\vec{B\Phi}\herm$ in \eqref{H1H0} as temporally white and Gaussian, as in \cite{Kelly:TAES:86,Gerlach:TSP:00,Kang:TAES:14,Aubry:TSP:13}.
In this case, the interference-plus-noise matrix
\begin{align}
\vec{N}\defn\vec{B\Phi}\herm+\vec{W}
\end{align}
is temporally white Gaussian with spatial covariance matrix $\vec{\Sigma}=\vec{R}+\nu\vec{I}_M$, where both $\vec{R}\geq 0$ and $\nu>0$ are unknown.
\textb{For now, we will model $\vec{R}$ using a fixed and known rank $N\leq M$.}
The $N=M$ case is reminiscent of Kelly \cite{Kelly:TAES:86}, and the $N<M$ case is reminiscent of Kang, Monga, and Rangaswamy \cite{Kang:TAES:14}.
\textb{The estimation of $N$ will be discussed in \secref{rankGS}.}

For a fixed rank $N$, the hypothesis test \eqref{H1H0} reduces to 
\begin{subequations}
\label{eq:H1H0a} 
\begin{align}
\mc{H}_1: \vec{Y}
&= \vec{h}\vec{s}\herm + 
\textb{
\vec{N} \text{~for~}
\vect(\vec{N})\sim\mc{CN}(\vec{0},\vec{I}_L\otimes\vec{\Sigma})
}
\label{eq:H1a} \\
\mc{H}_0: \vec{Y}
&= 
\textb{
\vec{N} \text{~for~}
\vect(\vec{N})\sim\mc{CN}(\vec{0},\vec{I}_L\otimes\vec{\Sigma})
} 
\label{eq:H0a},
\end{align}
\end{subequations}
where $\vec{h}$ and $\vec{\Sigma}\in\Sigset_N$ (defined in \eqref{SigsetN}) are unknown and $\vec{s}\sim p(\vec{s})$.
When $N=M$, note that $\vec{\Sigma}\in\Sigset_N$ reduces to $\vec{\Sigma}>0$.
The corresponding GLRT is
\begin{align}
\frac{\max_{\vec{h},\vec{\Sigma}\in\Sigset_N} p(\vec{Y}|\mc{H}_1;\vec{h},\vec{\Sigma})}
{\max_{\vec{\Sigma}\in\Sigset_N} p(\vec{Y}|\mc{H}_0;\vec{\Sigma})} 
\gtreqless \eta
\label{eq:glrta}.
\end{align}
As a consequence of $\vec{s}\sim p(\vec{s})$, the numerator likelihood in \eqref{glrta} differs from that in \eqref{glrt_kmr}, as detailed in the sequel.

\subsection{GLRT Denominator} \label{sec:GLRTdenGS}
\vspace{-0.5mm} 

For the denominator of \eqref{glrta}, equations \eqref{H0a} and \eqref{SigsetN} imply
\begin{align}
p(\vec{Y}|\mc{H}_0;\vec{\Sigma})
&= \frac{\exp(-\tr\{\vec{Y}\herm \vec{\Sigma}^{-1} \vec{Y}\})}
        {\pi^{ML} |\vec{\Sigma}|^L} \\
&= \left[\frac{\exp(-\tr\{\frac{1}{L}\vec{YY}\herm \vec{\Sigma}^{-1} \})}
        {\pi^M |\vec{\Sigma}|} \right]^L
\label{eq:pY0a} .
\end{align}
We first find the ML estimate $\hvec{\Sigma}_0$ of $\vec{\Sigma}\in\Sigset_N$ under $\mc{H}_0$.
When $N<M$, the results in \cite{Anderson:AMS:63} (see also \cite{Kang:TAES:14}) imply that
\begin{align}
\hvec{\Sigma}_0 &= \vec{V}_0\hvec{\Lambda}_0\vec{V}_0\herm, 
~~
\hvec{\Lambda}_0 = \Diag(\hat{\lambda}_{0,1},\dots,\hat{\lambda}_{0,M}) 
\label{eq:Sig0hata} ,
\end{align}
where $\{\hat{\lambda}_{0,m}\}_{m=1}^M$ follow the definition in \eqref{lamihata} with $i=0$.
That is, $\{\hat{\lambda}_{0,m}\}_{m=1}^M$ is a smoothed version of the eigenvalues $\{{\lambda}_{0,m}\}$ of the sample covariance matrix $\frac{1}{L}\vec{YY}\herm$ in decreasing order, where the smoothing averages the $M-N$ smallest eigenvalues to form the noise variance estimate $\hat{\nu}_0$, as in \eqref{vihata}.
When $N=M$, the results in \cite{Goodman:AMS:63} (see also \cite{Kelly:TAES:86}) imply that $\hat{\lambda}_{0,m}=\lambda_{0,m}~\forall m$.
In either case, the columns of $\vec{V}_0$ are the corresponding eigenvectors of the sample covariance matrix $\frac{1}{L}\vec{YY}\herm$.
Plugging \eqref{Sig0hata} into \eqref{pY0a}, taking the log, and rearranging gives
\begin{align}
\lefteqn{ 
\tfrac{1}{L}\ln p(\vec{Y}|\mc{H}_0;\hvec{\Sigma}_0) + M \ln\pi
}\nonumber\\
&= -\tr\big\{ \tfrac{1}{L}\vec{YY}\herm \hvec{\Sigma}_0^{-1} \big\} -\ln |\hvec{\Sigma}_0| \\
&=\sum_{m=1}^M \left(-\frac{\lambda_{0,m}}{\hat{\lambda}_{0,m}} - \ln \hat{\lambda}_{0,m} \right) \\
&=\sum_{m=1}^N \left(-1 - \ln \lambda_{0,m}\right)
 +\sum_{m=N+1}^M \left(-\frac{\lambda_{0,m}}{\hat{\nu}_0} - \ln \hat{\nu}_0 \right) .
\end{align}
Since $\sum_{m=N+1}^M\lambda_{0,m}=\hat{\nu}_0$, we have
\begin{align}
\lefteqn{ 
\tfrac{1}{L}\ln p(\vec{Y}|\mc{H}_0;\hvec{\Sigma}_0) + M \ln\pi
}\nonumber\\
&=-N - \sum_{m=1}^N \ln \lambda_{0,m}
 +(M-N)(-1-\ln \hat{\nu}_0) \\
&=-M - \sum_{m=1}^N \ln \lambda_{0,m}
 -(M-N) \ln \hat{\nu}_0 \\
&=-M - \sum_{m=1}^M \ln \hat{\lambda}_{0,m} 
\label{eq:logLike0a} .
\end{align}
When $N<M$, note that $\{\hat{\lambda}_{0,m}\}_{m=1}^M$ can be computed using only the $N$ principal eigenvalues of $\frac{1}{L}\vec{YY}\herm$, since
\begin{align}
\hat{\nu}_0
&=\frac{1}{M-N}\left( \tr\Big\{\frac{1}{L}\vec{YY}\herm\Big\}-\sum_{m=1}^N\lambda_{0,m} \right) .
\end{align}

\subsection{GLRT Numerator} \label{sec:GLRTnumGS}

For the numerator of \eqref{glrta}, $\vec{s}\sim p(\vec{s})$ and \eqref{H1a} imply
\begin{align}
\lefteqn{
p(\vec{Y}|\mc{H}_1;\vec{h},\vec{\Sigma}) 
= \int p(\vec{Y}|\vec{s},\mc{H}_1;\vec{h},\vec{\Sigma}) \,p(\vec{s}) \dif\vec{s} 
} \label{eq:pY1aa}\\
&= \int \frac{\exp(-\tr\{(\vec{Y}-\vec{h}\vec{s}\herm)\herm\vec{\Sigma}^{-1}(\vec{Y}-\vec{h}\vec{s}\herm)\})}{\pi^{ML} |\vec{\Sigma}|^L} p(\vec{s}) \dif\vec{s}
\label{eq:pY1a} .
\end{align}
Exact maximization of $p(\vec{Y}|\mc{H}_1;\vec{h},\vec{\Sigma})$ over $\vec{h}$ and $\vec{\Sigma}\in\Sigset_N$ appears to be intractable.
We thus propose to approximate the maximization by applying EM \cite{Dempster:JRSS:77} with hidden data $\vec{s}$. 
This implies that we iterate the following over $t=0,1,2,\dots$:
\begin{align}
\lefteqn{\big(\hvec{h}\of{t+1},\hvec{\Sigma}_1\of{t+1}\big)}\label{eq:EMa}\\
&= \argmax_{\vec{h}\in\Complex^M,\vec{\Sigma}\in\Sigset_N} \E\big\{\ln p(\vec{Y},\vec{s}|\mc{H}_1;\vec{h},\vec{\Sigma}) \,\big|\, \vec{Y}; \hvec{h}\of{t},\hvec{\Sigma}_1\of{t}\big\}
\nonumber 
\end{align}
The EM algorithm is guaranteed to converge to a local maxima or saddle point of the likelihood \eqref{pY1aa} \cite{Wu:AS:83}.
Furthermore, at each iteration $t$, the EM-approximated log-likelihood increases and lower bounds the true log-likelihood \cite{Neal:Jordan:98}.

\textb{Because $\vec{s}$ is statistically independent of $\vec{h}$ and $\vec{\Sigma}$,
we have 
$\ln p(\vec{Y},\vec{s}|\mc{H}_1;\vec{h},\vec{\Sigma}) = 
\ln p(\vec{Y}|\vec{s},\mc{H}_1;\vec{h},\vec{\Sigma})+ \ln p(\vec{s})$,
which allows us to rewrite \eqref{EMa} as} 
\begin{align}
\lefteqn{
\argmax_{\vec{h}\in\Complex^M,\vec{\Sigma}_1\in\Sigset_N} 
\E\big\{\ln p(\vec{Y}|\vec{s},\mc{H}_1;\vec{h},\vec{\Sigma}) \,\big|\, \vec{Y}; \hvec{h}\of{t},\hvec{\Sigma}_1\of{t}\big\} 
}\\
&= \argmin_{\vec{h}\in\Complex^M,\vec{\Sigma}\in\Sigset_N} 
\int \Big[
\tr\{(\vec{Y}-\vec{h}\vec{s}\herm)\herm\vec{\Sigma}^{-1}(\vec{Y}-\vec{h}\vec{s}\herm)\}
\nonumber\\&\hspace{25mm}
+ \ln |\vec{\Sigma}|^L \Big]  
p(\vec{s}|\vec{Y};\hvec{h}\of{t},\hvec{\Sigma}_1\of{t}) \dif\vec{s} 
\label{eq:Mstep1a}.
\end{align}
We first perform the minimization in \eqref{Mstep1a} over $\vec{h}$.
Since 
\begin{align}
\lefteqn{
\tr\{(\vec{Y}-\vec{h}\vec{s}\herm)\herm\vec{\Sigma}^{-1}(\vec{Y}-\vec{h}\vec{s}\herm)\}} \nonumber\\
&= 
\tr\{\vec{Y}\herm\vec{\Sigma}^{-1}\vec{Y}\}
-\vec{h}\herm\vec{\Sigma}^{-1}\vec{Ys}
-\vec{s}\herm\vec{Y}\herm\vec{\Sigma}^{-1}\vec{h}
\nonumber\\&\quad
+\vec{h}\herm\vec{\Sigma}^{-1}\vec{h}\|\vec{s}\|^2 ,
\end{align}
the gradient of the cost in \eqref{Mstep1a} w.r.t.\ $\vec{h}$ equals
\begin{align}
2\int 
\Big[ \vec{\Sigma}^{-1}\vec{h}\|\vec{s}\|^2 -\vec{\Sigma}^{-1}\vec{Ys} \Big]
p(\vec{s}|\vec{Y};\hvec{h}\of{t},\hvec{\Sigma}_1\of{t}) \dif\vec{s} ,
\end{align}
and this gradient is set to zero by 
\begin{align}
\hvec{h}\of{t+1} 
&= \frac{\vec{Y}\E\{\vec{s}|\vec{Y};\hvec{h}\of{t},\hvec{\Sigma}_1\of{t}\}}
   {\E\{\|\vec{s}\|^2 |\vec{Y};\hvec{h}\of{t},\hvec{\Sigma}_1\of{t}\} }
= \frac{\vec{Y}\hvec{s}\of{t}}{E\of{t}} ,
\end{align}
which uses the notation
\begin{align}
\hvec{s}\of{t}
&\defn \E\{\vec{s}|\vec{Y};\hvec{h}\of{t},\hvec{\Sigma}_1\of{t}\} 
\label{eq:shat} \\
E\of{t} 
&\defn \E\{\|\vec{s}\|^2|\vec{Y};\hvec{h}\of{t},\hvec{\Sigma}_1\of{t}\} 
\label{eq:E} .
\end{align}
\vspace{-5mm} 

Setting $\vec{h}=\hvec{h}\of{t+1}$ in \eqref{Mstep1a}, we obtain the cost that must be minimized over $\vec{\Sigma}\in\Sigset_N$:
\begin{align}
&
\tr\{\vec{Y}\herm\vec{\Sigma}^{-1}\vec{Y}\}
-\hvec{h}\ofH{t+1}\vec{\Sigma}^{-1}\vec{Y}\hvec{s}\of{t}
-\hvec{s}\ofH{t}\vec{Y}\herm\vec{\Sigma}^{-1}\hvec{h}\of{t+1} 
\nonumber\\&\quad
+\hvec{h}\ofH{t+1}\vec{\Sigma}^{-1}\hvec{h}\of{t+1} E\of{t}
+ \ln |\vec{\Sigma}|^L
\nonumber\\
&= \tr\{\vec{Y}\herm\vec{\Sigma}^{-1}\vec{Y}\}
-\frac{\hvec{s}\ofH{t}\vec{Y}\herm\vec{\Sigma}^{-1}\vec{Y}\hvec{s}\of{t}}{E\of{t}}
+ \ln |\vec{\Sigma}|^L \\
&= \tr\big\{
\vec{Y} \tvec{P}_{\hvec{s}\of{t}}^\perp \vec{Y}\herm \vec{\Sigma}^{-1}
\big\} + \ln |\vec{\Sigma}|^L
\label{eq:Mstep2a},
\end{align}
where 
\begin{align}
\tvec{P}_{\hvec{s}\of{t}}^\perp 
&\defn \vec{I}_L - \frac{\hvec{s}\of{t}\hvec{s}\ofH{t}}{E\of{t}}  
\label{eq:Ptilsperp}\\
&= \vec{P}_{\hvec{s}\of{t}}^\perp + \vec{P}_{\hvec{s}\of{t}} - \frac{\hvec{s}\of{t}\hvec{s}\ofH{t}}{\|\hvec{s}\|^2} \frac{\|\hvec{s}\|^2}{E\of{t}} 
   \end{align} 
   \begin{align} 
&= \vec{P}_{\hvec{s}\of{t}}^\perp + \vec{P}_{\hvec{s}\of{t}} \frac{E\of{t}-\|\hvec{s}\|^2}{E\of{t}} \\
&= \vec{P}_{\hvec{s}\of{t}}^\perp + \vec{P}_{\hvec{s}\of{t}} \frac{\tr[\Cov\{\vec{s}|\vec{Y};\hvec{h}\of{t},\vec{\Sigma}\of{t}\}]}{E\of{t}} .
\end{align}
Note that $\tvec{P}_{\hvec{s}\of{t}}^\perp$ is a regularized version of the projection matrix $\vec{P}_{\hvec{s}\of{t}}^\perp$ that equals $\vec{P}_{\hvec{s}\of{t}}^\perp$ when $\vec{s}$ is completely known.
In general, however, $\tvec{P}_{\hvec{s}\of{t}}^\perp$ is not a projection matrix.
Minimizing \eqref{Mstep2a} over $\vec{\Sigma}\in\Sigset_N$ is equivalent to maximizing 
\begin{align}
\frac{ \exp(- \tr\{\vec{Y} \tvec{P}_{\hvec{s}\of{t}}^\perp \vec{Y}\herm \vec{\Sigma}^{-1}\}) }
{ \pi^{ML} |\vec{\Sigma}|^L } .
\end{align}
As with \eqref{pY0a}, when $N<M$, the results in \cite{Anderson:AMS:63} imply 
\begin{align}
\hvec{\Sigma}_1\of{t+1} 
&= \vec{V}_1\of{t+1}\hvec{\Lambda}_1\of{t+1}\vec{V}_1\ofH{t+1}
\label{eq:Sig1hata},\\
\hvec{\Lambda}_1\of{t+1} 
&= \Diag(\hat{\lambda}_{1,1}\of{t+1},\dots,\hat{\lambda}_{1,M}\of{t+1})\\
\hat{\lambda}_{1,m}\of{t+1} 
&\textb{\defn} \begin{cases}
\lambda_{1,m}\of{t+1} & m=1,\dots,N \\
\hat{\nu}_1\of{t+1} & m=N+1,\dots,M \\
\end{cases}
\label{eq:lam1hata}\\
\hat{\nu}_1\of{t+1} &\defn \frac{1}{M-N}\sum_{m=N+1}^M \lambda_{1,m}\of{t+1} 
\label{eq:v1hata},
\end{align}
where $\{\lambda_{1,m}\of{t+1}\}_{m=1}^M$ are the eigenvalues of the matrix $\frac{1}{L}\vec{Y}\tvec{P}_{\hvec{s}\of{t}}^\perp\vec{Y}\herm$ in decreasing order, and the columns of $\vec{V}_1\of{t+1}$ are the corresponding eigenvectors.
When $N=M$, we have that $\hat{\lambda}_{1,m}=\lambda_{1,m}~\forall m$.


We have thus derived the EM procedure that iteratively lower bounds \cite{Neal:Jordan:98} the numerator of \eqref{glrta} under a generic signal prior $p(\vec{s})$.

\subsection{EM Update under an Independent Prior} \label{sec:EMupdate}

\textb{The EM updates of $\hvec{s}\of{t}$ and $E\of{t}$ in \eqref{shat}-\eqref{E} compute the conditional mean (or, equivalently, the MMSE estimate \cite{Scharf:Book:91}) of $\vec{s}$ and $\|\vec{s}\|^2$, respectively, given the measurements $\vec{Y}$ in \eqref{H1H0a} under the model $\vec{h}=\hvec{h}\of{t}$ and $\vec{\Sigma}=\hvec{\Sigma}\of{t}_1$.}
For any independent prior, as in \eqref{s_indep}, we can MMSE-estimate the symbols one at a time from the measurement equation
\begin{align}
\vec{y}_l
&= \hvec{h}\of{t}s_l^* + \mc{CN}(\vec{0},\hvec{\Sigma}_1\of{t}) .
\end{align}
From $\vec{y}_l$, we obtain a sufficient statistic \cite{Scharf:Book:91} for the estimation of $s_l$ by spatially whitening the measurements via 
\begin{align}
\tvec{y}_l\of{t}
\defn 
(\hvec{\Sigma}_1\of{t})^{-\frac{1}{2}}\vec{y}_l
&= (\hvec{\Sigma}_1\of{t})^{-\frac{1}{2}} \hvec{h}\of{t}s_l^* + \mc{CN}(\vec{0},\vec{I}) 
\end{align}
and then matched filtering via
\begin{align}
\tilde{r}_l\of{t}
\defn \hvec{h}\ofH{t}(\hvec{\Sigma}_1\of{t})^{-\frac{1}{2}}\tvec{y}_l\of{t}
&= \xi\of{t} s_l^* + \mc{CN}(\vec{0},\xi\of{t}) ,
\end{align}
where 
\begin{align}
\xi\of{t}
&\defn \hvec{h}\ofH{t}(\hvec{\Sigma}_1\of{t})^{-1}\hvec{h}\of{t} 
\label{eq:xi} .
\end{align}
We find it more convenient to work with the normalized and conjugated statistic
\begin{align}
r_l\of{t} 
\defn \frac{\tilde{r}_l\ofc{t}}{\xi\of{t}} 
&= s_l + \mc{CN}\bigg(0,\frac{1}{\xi\of{t}}\bigg) 
\label{eq:rl},
\end{align}
which is a Gaussian-noise-corrupted version of the true symbol $s_l$, with noise precision $\xi\of{t}$.

The computation of the MMSE estimate $\hat{s}_l$ from $r_l\of{t}$ depends on the prior $p(s_l)$.
For the Gaussian prior $p(s_l)=\mc{CN}(s_l;\mu_l,v_l)$, we have the posterior mean and variance \cite{Scharf:Book:91}
\begin{align}
\hat{s}_l
&= \mu_l + \frac{v_l}{v_l + 1/\xi\of{t}}(r_l\of{t}-\mu_l) 
\label{eq:shatl_gauss} \\
\hat{v}_l
&= \frac{1}{\xi\of{t} + 1/v_l} ,
\end{align}
which from \eqref{E} implies
\begin{align}
E\of{t}
&= \sum_{l=1}^L \E\{|s_l|^2 \,|\, \vec{Y};\hvec{h}\of{t},\hvec{\Sigma}_1\of{t}\}
= \sum_{l=1}^L \left( |\hat{s_l}|^2 + \hat{v}_l \right)
\label{eq:E_gauss} .
\end{align}
For the discrete prior 
$p(s_l) = \sum_{k=1}^{K_l} \omega_{lk} \delta(s_l-d_{lk})$, 
with alphabet $\mc{A}_l=\{d_{lk}\}_{k=1}^{K_l}$ and prior symbol probabilities $\omega_{lk}\geq 0$ (such that $\sum_{k=1}^{K_l} \omega_{lk} = 1~\forall l$), it is straightforward to show that the posterior density is
\begin{align}
p(s_l|r_l\of{t})
&= \sum_{k=1}^{K_l} \hat{\omega}_{lk}\of{t} \delta(s_l-d_{lk}) \\
\hat{\omega}_{lk}\of{t}
&\defn \frac{\omega_{lk} \mc{CN}(d_{lk};r_l\of{t},1/\xi\of{t})}
        {\sum_{k'=1}^{K_l} \omega_{lk'} \mc{CN}(d_{lk'};r_l\of{t},1/\xi\of{t})} ,
\end{align}
and thus the posterior mean and second moment are
\begin{align}
\hat{s}_l\of{t}
&= \sum_{k=1}^{K_l} \hat{\omega}_{lk}\of{t} d_{lk} 
\label{eq:shatl_discrete}\\
\E\{|\hat{s}_l\of{t}|^2 \,|\, \vec{Y};\hvec{h}\of{t},\hvec{\Sigma}_1\of{t}\}
&= \sum_{k=1}^{K_l} \hat{\omega}_{lk}\of{t} |d_{lk}|^2 ,
\end{align}
which from \eqref{E} implies
\begin{align}
E\of{t}
&= \sum_{l=1}^L \sum_{k=1}^{K_l} \hat{\omega}_{lk}\of{t} |d_{lk}|^2 
\label{eq:E_discrete} .
\end{align}
This EM update procedure is summarized in \algref{GS}.

\begin{algorithm}[t]
\footnotesize
\caption{EM update under white Gaussian interference}
\label{alg:GS}
\begin{algorithmic}[1]
\REQUIRE{
    Data $\vec{Y}\in\Complex^{M\times L}$,
    signal prior $p(\vec{s})=\prod_{l=1}^L p_l(s_l)$.
}
\smallskip
\STATE
Initialize $\hvec{s}$ and $E>0$ (see \secref{initGS})
\smallskip
\REPEAT
\smallskip
  \STATE 
    $\hvec{h} \leftarrow \frac{1}{E}\vec{Y}\hvec{s}$ 
    \label{line:h}
  \STATE 
    $\hvec{\Sigma}_1 \leftarrow \frac{1}{L}\vec{YY}\herm 
                - \frac{E}{L}\hvec{h}\hvec{h}\herm$
    \label{line:Sig}
  \STATE
    Estimate interference rank $N$ (see \secref{rankGS}).
  \IF{$N=0$}
  \STATE
    $\hat{\nu}_1 \leftarrow \frac{1}{M}\tr(\hvec{\Sigma}_1)$
  \STATE
    $\vec{g} \leftarrow \frac{1}{\hat{\nu}_1}\hvec{h}$
  \ELSIF{$N=M$}
  \STATE
    $\vec{g} \leftarrow \hvec{\Sigma}_1^{-1} \hvec{h}$
    \label{line:inv}
  \ELSE
  \STATE 
    $\big\{\ovec{V}_1,\ovec{\Lambda}_1\big\}
     \leftarrow \text{principal\_eigs}( \hvec{\Sigma}_1,N )$
    \label{line:eigs}
  \STATE 
    $\hat{\nu}_1 \leftarrow \frac{1}{M-N}\big( \tr(\hvec{\Sigma}_1)
                - \tr\{\ovec{\Lambda}_1\} \big)$
  \STATE 
    $\vec{g} \leftarrow \frac{1}{\hat{\nu}_1}\hvec{h}
                + \ovec{V}_1
                \big(\ovec{\Lambda}_1^{-1}-\frac{1}{\hat{\nu}_1}\vec{I}_N\big)
                \ovec{V}_1\herm \hvec{h}$
  \ENDIF
  \STATE 
    $\xi \leftarrow \hvec{h}\herm\vec{g}$
    \label{line:xi}
  \STATE
    $\vec{r} \leftarrow \frac{1}{\xi}\vec{Y}\herm\vec{g}
     \text{~where~} \vec{r} \sim \mc{CN}(\vec{s},\vec{I}/\xi)$
    \label{line:r}
  \STATE 
    $\hat{s}_l \leftarrow \E\{s_l | r_l; \xi\}~\forall l=1,\dots,L$
    \label{line:shat}
  \STATE 
    $E \leftarrow \sum_{l=1}^L \E\{|s_l|^2 | r_l; \xi\}$
    \label{line:E}
\smallskip
\UNTIL{Terminated}
\end{algorithmic}
\end{algorithm}

\subsection{Fast Implementation of \Algref{GS}} \label{sec:fastGS}

The implementation complexity of \algref{GS} is dominated by the eigenvalue decomposition in \lineref{eigs}, which consumes $O(M^3)$ operations per EM iteration.
We now describe how the complexity of this step can be reduced.
Recall that 
\begin{align}
\tfrac{1}{L}\vec{YY}\herm
&= \vec{V}_0\vec{\Lambda}_0\vec{V}_0\herm
\label{eq:Sig0} ,
\end{align}
as described after \eqref{Sig0hata}.
Thus $\hvec{\Sigma}_1$ in \lineref{Sig} takes the form 
\begin{align}
\hvec{\Sigma}_1
&= \vec{V}_0\vec{\Lambda}_0\vec{V}_0\herm - \tfrac{E}{L}\hvec{h}\hvec{h}\herm\\
&= \vec{V}_0\big( \vec{\Lambda}_0 - \tvec{h}\tvec{h}\herm
   \big)\vec{V}_0\herm
\end{align}
using the definition 
\begin{align}
\tvec{h} 
\defn \sqrt{\tfrac{E}{L}}\vec{V}_0\herm\hvec{h} .
\end{align}
The key idea is that the eigen-decomposition of $\vec{\Lambda}_0 - \tvec{h}\tvec{h}\herm$ can be computed in a fast manner due to its diagonal-plus-rank-one structure
\cite{Gu:JMAA:94}.

We now provide some details.
First, define $R\defn \rank(\tfrac{1}{L}\vec{YY}\herm)$, where $R\leq M$.
Without loss of generality, suppose that 
$\vec{V}_0$ has $R$ columns and that
$\vec{\Lambda}_0\in\Real^{R\times R}$, and assume that these quantities have been computed before the start of the EM iterations.
Then $\tvec{h}$ can be computed in $O(MR)$ operations,
the eigen-decomposition $\vec{Q}\vec{\Lambda}_1\vec{Q}\herm = \vec{\Lambda}_0 - \tvec{h}\tvec{h}\herm$ can be computed in $O(R^2)$ operations 
\cite{Gu:JMAA:94},
and the eigenvectors $\vec{V}_1=\vec{V}_0\vec{Q}$ of $\hvec{\Sigma}_1$ can be computed in $O(MR^2)$ operations.
Since only the $N$ principal eigenvectors are needed for \lineref{eigs}, the latter reduces to $O(MRN)$ operations.

\subsection{Evaluation of the GLRT}

We now describe what remains of the GLRT.
Let us denote the final EM-based estimates of $\vec{s}$, $\vec{h}$, and $\vec{\Sigma}$ under $\mc{H}_1$ as $\hvec{s}$, $\hvec{h}$, and $\hvec{\Sigma}_1$, respectively.
Notice that 
\begin{align}
\lefteqn{ 
\tfrac{1}{L} \ln p(\vec{Y}|\mc{H}_1;\hvec{h},\hvec{\Sigma}_1) 
}\nonumber\\
&= -\tr\big\{ \tfrac{1}{L}\vec{Y}\tvec{P}_{\hvec{s}}^\perp\vec{Y}\herm \hvec{\Sigma}_1^{-1} \big\} -\ln |\hvec{\Sigma}_1| - M \ln \pi 
\\
&=-M - \sum_{m=1}^M \ln \hat{\lambda}_{1,m} -M\ln\pi 
\label{eq:logLike1a} ,
\end{align}
following steps similar to \eqref{logLike0a}.
Recalling \eqref{glrta}, the log-domain GLRT is obtained by subtracting \eqref{logLike0a} from \eqref{logLike1a}, yielding
\begin{align}
\sum_{m=1}^M \ln \frac{\hat{\lambda}_{0,m}}{\hat{\lambda}_{1,m}}
\gtreqless \eta' .
\end{align}
When $N<M$, this test can be simplified by recalling that the smallest $M-N$ eigenvalues in $\{\hat{\lambda}_{i,m}\}$ equal $\hat{\nu}_i$ for $i=0,1$.
In this case, the log-domain GLRT reduces to
\begin{align}
\sum_{m=1}^N \ln \frac{\hat{\lambda}_{0,m}}{\hat{\lambda}_{1,m}}
+ (M-N) \ln \frac{\hat{\nu}_0}{\hat{\nu}_1}
\gtreqless \eta'
\label{eq:logGlrta} .
\end{align}

\textb{Although the proposed GLRT is not CFAR \cite{Scharf:Book:91}, neither is the simpler Kang/Monga/Rangaswamy (KMR) \cite{Kang:TAES:14} detector that results in the special case where $\vec{s}$ is known (i.e., $p(\vec{s})$ is a point mass).
So, to set the threshold $\eta'$ in practice, one could run experiments or simulations to provide histograms of the test statistic under $\mc{H}_0$ and $\mc{H}_1$, and then choose the value of $\eta'$ that yields the desired balance between miss rate and false-alarm rate.
}

\subsection{Relation to Forsythe's Iterative Method} \label{sec:Forsythe}

We now connect the above method to Forsythe's iterative scheme in \cite[p.110]{Forsythe:LLJ:97}, which assumes full-rank interference (i.e., $N=M$) and positive definite sample covariance, i.e., $\frac{1}{L}\vec{YY}\herm>0$.
To make this connection, we \textb{find it convenient\footnote{%
\textb{We can transform from $\vec{Y}$ to $\uvec{Y}$ and back without loss of generality because the transformation is invertible.}} 
to work with}
the spatially whitened measurements
\begin{align}
\uvec{Y}
&\defn (\tfrac{1}{L}\vec{YY}\herm)^{-\frac{1}{2}} \vec{Y} 
\label{eq:whiten}.
\end{align}
Writing lines~\ref{line:h}, \ref{line:Sig}, \ref{line:xi}, and \ref{line:r} of \algref{GS} in terms of the whitened quantities 
$\underline{\hvec{h}}\defn(\tfrac{1}{L}\vec{YY}\herm)^{-\frac{1}{2}} \hvec{h}$
and
$\underline{\hvec{\Sigma}}_1\defn (\tfrac{1}{L}\vec{YY}\herm)^{-\frac{1}{2}} \hvec{\Sigma}_1 (\tfrac{1}{L}\vec{YY}\herm)^{-\frac{1}{2}}$
gives
\begin{align}
\underline{\hvec{h}}
&= \tfrac{1}{E} \uvec{Y}\hvec{s} 
\label{eq:hwhite}\\
\underline{\hvec{\Sigma}}_1
&= \tfrac{1}{L} \uvec{Y}\uvec{Y}\herm - \tfrac{E}{L}\underline{\hvec{h}}\underline{\hvec{h}}\herm 
\label{eq:Sig1white}\\
\xi
&= \underline{\hvec{h}}\herm\underline{\hvec{\Sigma}}_1^{-1}\underline{\hvec{h}}
\label{eq:xiwhite}\\
\vec{r}
&= \xi^{-1}\uvec{Y}\herm\underline{\hvec{\Sigma}}_1^{-1}\underline{\hvec{h}}
\label{eq:rwhite}.
\end{align}
From the construction of $\uvec{Y}$ and the assumption $\frac{1}{L}\vec{YY}\herm>0$, we have $\tfrac{1}{L} \uvec{Y}\uvec{Y}\herm=\vec{I}_M$. 
Thus, applying the matrix inversion lemma to \eqref{Sig1white} gives
\begin{align}
\underline{\hvec{\Sigma}}_1^{-1} 
&= \vec{I}_M - \left(\tfrac{L}{E}+\|\underline{\hvec{h}}\|^2\right)^{-1} \underline{\hvec{h}}\underline{\hvec{h}}\herm 
\label{eq:mil} .
\end{align}
Plugging \eqref{mil} into \eqref{rwhite}, we obtain 
\begin{align}
\vec{r}
&= \frac{\uvec{Y}\herm\underline{\hvec{h}}}{\|\underline{\hvec{h}}\|^2} 
= \frac{\uvec{Y}\herm\uvec{Y}\hvec{s} E}{\|\uvec{Y}\hvec{s}\|^2} 
\label{eq:rwhite_2} ,
\end{align}
which can be expressed in terms of unwhitened quantities as
\begin{align}
\vec{r}
&= \frac{\vec{Y}\herm (\frac{1}{L}\vec{Y}\vec{Y}\herm)^{-1}\vec{Y}\hvec{s} E}{\|(\frac{1}{L}\vec{YY}\herm)^{-\frac{1}{2}}\vec{Y}\hvec{s}\|^2} 
= \frac{\vec{Y}\herm (\vec{Y}\vec{Y}\herm)^{-1}\vec{Y}\hvec{s} E}{\hvec{s}\herm\vec{Y}\herm(\vec{YY}\herm)^{-1}\vec{Y}\hvec{s}} \\
&= \vec{Y}\herm 
\underbrace{ 
(\vec{Y}\vec{Y}\herm)^{-1}\vec{Y}\hvec{s} \frac{E}{\|\vec{P}_{\vec{Y}\herm}\hvec{s}\|^2} 
}_{\displaystyle \defn \vec{w}} 
.
\end{align}

\Algref{GS} prescribes the use of the ``soft'' symbol estimate 
$\hvec{s}=\E\{\vec{s}|\vec{r};\xi\}$
and the soft squared-norm estimate $E=\E\{\|\vec{s}\|^2\,|\vec{r};\xi\}$
in lines~\ref{line:shat}-\ref{line:E}.
If we replaced these soft estimates with ``hard'' estimates, i.e., the ML
estimate $\hvec{s}\ml=\argmin_{\vec{s}\in\mc{A}^L}\|\vec{r}-\vec{s}\|^2$ and 
its squared-norm $E\ml=\|\hvec{s}\ml\|^2$, then \algref{GS} would become
\begin{align}
\vec{w}
&\leftarrow (\vec{Y}\vec{Y}\herm)^{-1}\vec{Y}\hvec{s}\ml \frac{\|\hvec{s}\ml\|^2}{\|\vec{P}_{\vec{Y}\herm}\hvec{s}\ml\|^2} \\
\vec{r}
&\leftarrow \vec{Y}\herm\vec{w} \\
\hvec{s}\ml
&\leftarrow \argmin_{\vec{s}\in\mc{A}^L}\|\vec{r}-\vec{s}\|^2 ,
\end{align}
which is precisely Forsythe's iterative method from \cite[p.110]{Forsythe:LLJ:97}.
There, $\vec{w}$ is interpreted as a least-squares (LS) beamformer.
We have thus shown that \algref{GS} under fixed rank $N=M$ is a soft version of Forsythe's iterative method.
As we will show later, the soft nature of \algref{GS} helps to prevent error propagation.

\subsection{Estimating the Interference Rank \texorpdfstring{$N$}{N}} \label{sec:rankGS}

We now consider estimation of the interference rank $N=\rank(\vec{R})$.
For this, we adopt the standard information-theoretic model-order selection approach described in, e.g., \cite{Wax:TASSP:86,Stoica:SPM:04}, which specifies
\begin{align}
\hat{N}
&=\argmax_{N=0,\dots,N_{\max}} \ln p(\vec{Y}|\mc{H}_1;\hvec{\Theta}_N) - J(D(N))
\label{eq:mos} ,
\end{align}
where 
$J(\cdot)$ is a penalty function,
$\hvec{\Theta}_N$ is the ML parameter estimate under rank hypothesis $N$,
and $D(N)$ is the degrees-of-freedom (DoF) in the parameters $\vec{\Theta}_N$.
Common choices of $J(\cdot)$ include
\begin{align}
J(D)
&= \begin{cases}
D & \text{Akaike's Information Criterion (AIC)} \\
\frac{TD}{T-D-1} & \text{Corrected AIC (AICc)} \\
\frac{D}{2}\ln T & \text{Bayesian Information Criterion (BIC)} \\
G D & \text{Generalized Information Criterion (GIC)} \\
\end{cases}
\label{eq:GIC}
\end{align}
where $T$ is the number of real-valued measurements and $G>0$ is a tunable gain.
The above BIC rule is the same as that which results from Rissanen's Minimum Description Length (MDL) criterion $T$ (see \cite{Wax:TASSP:86}).

For \algref{GS}, we have $T=2ML$ and 
\begin{align}
\vec{\Theta}_N = \{\vec{h},\vec{\Sigma}\} 
\text{~for~}
\vec{h}\in\Complex^M
\text{~and~}
\vec{\Sigma}\in\Sigset_N ,
\end{align}
with $\Sigset_N$ defined in \eqref{SigsetN}.
Here, the DoF in $\vec{h}$ equals $2M$ and
the DoF in $\vec{\Sigma}$ equals $(2M-N)N + 1$, since
the DoF in a $M\times M$ rank-$N$ Hermitian matrix $\vec{R}$ is $(2M-N)N$ 
and the DoF in the noise variance $\nu$ is $1$.
In summary, $D(N)=(2M-N)N + 2M + 1$.
For our numerical experiments, we used GIC with $G=10$. 

\subsection{EM Initialization} \label{sec:initGS}

\textb{The EM algorithm is guaranteed to converge to a local maxima or saddle point of the likelihood \eqref{pY1aa} \cite{Wu:AS:83} under mild technical conditions.}
With a multi-modal likelihood, the initialization of $(\hvec{s},E)$ affects the quality of the final EM estimate.
Below, we propose an initialization assuming the training/data structure in \eqref{std}.
That is, $\vec{Y}=\mat{\vec{Y}\train&\vec{Y}\data}$ 
with 
\begin{align}
\vec{Y}\train &=\vec{h}\vec{s}\train\herm+\vec{N}\train,
~\vect(\vec{N}\train)\sim\mc{CN}(\vec{0},\vec{I}_Q\otimes\vec{\Sigma}) 
\label{eq:Yt} 
\\
\vec{Y}\data
&=\vec{h}\vec{s}\data\herm+\vec{N}\data,
~\vect(\vec{N}\data)\sim\mc{CN}(\vec{0},\vec{I}_{L-Q}\otimes\vec{\Sigma}) 
\label{eq:Yd} ,
\end{align}
and $\vec{s}=[\vec{s}\train\herm,\vec{s}\data\herm]\herm$.
Essentially, we would like to estimate the random vector $\vec{s}\data\sim \prod_{l=Q+1}^L p_l(s_l)$ from measurements $\vec{Y}$ under known $\vec{s}\train$ but unknown $\vec{s}\data,\vec{h},\vec{\Sigma},\vec{N}$.

Recall that the whitened matched-filter (WMF) outputs
\begin{align}
r_l 
&\defn \vec{y}_l\herm\vec{\Sigma}^{-1}\vec{h} 
\text{~for~} l\in\{Q+1,\dots,L\} 
\label{eq:wmf} 
\end{align}
are sufficient statistics \cite{Scharf:Book:91} for estimating $\vec{s}\data$.
Because $\vec{\Sigma}$ and $\vec{h}$ are unknown in our case, we propose to estimate them from the training data $\vec{Y}\train$ and use the results to compute approximate-WMF outputs of the form
\begin{align}
\hat{r}_l
&\defn \vec{y}_l\herm\hvec{\Sigma}\train^{-1}\hvec{h}\train 
\label{eq:wmf_hat} .
\end{align}
With appropriate scaling $\beta\in\Complex$, we get an unbiased statistic
\begin{align}
\beta\hat{r}_l
&\approx s_l + \mc{CN}(0,1/\hat{\xi})
\text{~for~} l\in\{Q+1,\dots,L\} 
\label{eq:wmf_unbiased} 
\end{align}
that can be converted to MMSE symbol estimates $\hat{s}_l$ via \eqref{shatl_gauss} or \eqref{shatl_discrete}, which are suitable for EM initialization. 
Likewise, the initialization of $E$ can be computed from \eqref{E_gauss} or \eqref{E_discrete}.

As for the choice of $(\hvec{\Sigma}\train,\hvec{h}\train)$ in \eqref{wmf_hat}, one possibility is the joint ML estimate of $\vec{\Sigma}\in\Sigset_N$ and $\vec{h}\in\Complex^M$ from the training $\vec{Y}\train$, assuming known interference rank $N$.
The arguments in \secref{GLRTnumGS} reveal that these joint-ML estimates equal
\begin{align}
\hvec{h}\train 
&\defn \frac{\vec{Y}\train \vec{s}\train}{\|\vec{s}\train\|^2} 
\label{eq:htrain}\\
\hvec{\Sigma}\train\of{N}
&\defn \vec{V}\train 
        \Diag\big( \hat{\lambda}\trainone\of{N},\dots,\hat{\lambda}\trainM\of{N} ) 
        \vec{V}\train\herm ,
\end{align}
where
\begin{align}
\hat{\lambda}\trainm\of{N}
&\defn \begin{cases}
\lambda\trainm & m=1,\dots,N \\
\frac{1}{M-N} \sum_{m'=N+1}^M \lambda\trainmprime 
& m=N\!+\!1,\dots,M, \\
\end{cases} 
\label{eq:lam1init} 
\end{align}
such that $\{\lambda\trainm\}_{m=1}^M$ are the eigenvalues of the sample covariance matrix 
\begin{align}
\hvec{\Sigma}\train 
&\defn \frac{1}{Q}\vec{Y}\train\vec{P}_{\vec{s}\train}^\perp\vec{Y}\train\herm
\label{eq:Sigtrain}
\end{align}
in decreasing order and $\vec{V}\train$ contains the eigenvectors.
When the interference rank $N$ is unknown, the methods in \secref{rankGS} can be used to estimate $N$ from $\vec{Y}\train$.
However, the estimation of the unbiasing gain $\beta$ and the precision $\hat{\xi}$ in \eqref{wmf_unbiased} remain challenging.

Instead of rank-$N$ covariance estimation, we propose to use a regularized estimate of the form \cite{Hoffbeck:TPAMI:96}
\begin{align}
\hvec{\Sigma}\train\of{\alpha} 
&= (1-\alpha)\hvec{\Sigma}\train + \alpha c\vec{I}_M, ~\alpha\in(0,1] ,
\end{align}
with $\hvec{\Sigma}\train$ from \eqref{Sigtrain} and $c \defn \tr(\hvec{\Sigma}\train)/M$.
Since the goal of regularization is robust estimation under possibly few training samples $Q$,
we propose to choose $\alpha$ to maximize (post-unbiased) precision $\hat{\xi}$, where the precision is estimated via leave-one-out cross-validation (LOOCV) \cite{Arlot:SS:10} on the training data.
Our LOOCV approach is similar to the ``SEO'' scheme from \cite{Tong:TSP:16} but targets minimum-variance unbiased estimation rather than MMSE estimation and, more significantly, handles non-white interference.
Details are provided below.

We first define the leave-one-out training quantities
$\vec{Y}\notl\defn[\vec{y}_1,\dots,\vec{y}_{l-1},\vec{y}_{l+1},\dots,\vec{y}_Q]$ and
$\vec{s}\notl\defn[s_1,\dots,s_{l-1},s_{l+1},\dots,s_Q]\tran$.
From these, we construct the ML $\vec{h}$-estimate and $\alpha$-regularized sample covariance 
\begin{align}
\hvec{h}\notl 
&\defn \frac{\vec{Y}\notl \vec{s}\notl}{\|\vec{s}\notl\|^2} \\
\hvec{\Sigma}\of{\alpha}\notl 
&\defn (1-\alpha)\frac{1}{Q-1}\vec{Y}\notl\vec{P}_{\vec{s}\notl}^\perp\vec{Y}\notl\herm + \alpha c\vec{I}_M ,
\end{align}
which can be used to form the out-of-sample estimate
\begin{align}
\hat{r}_l\of{\alpha}
&\defn \vec{y}_l\herm\big(\hvec{\Sigma}\notl\of{\alpha}\big)^{-1}\hvec{h}\notl 
\label{eq:wmf_alpha}.
\end{align}
It can be shown that 
\begin{align}
\hvec{h}\notl
&= \hvec{h}\train - \frac{s_l}{\|\vec{s}\notl\|^2}\hvec{n}_l 
\label{eq:hnotl}\\
\text{for~~}
\hvec{n}_l 
&\defn \vec{y}_l - \hvec{h}\train s_l^* .
\end{align}
Also, using the matrix inversion lemma, it can be shown that
\begin{align}
\big(\hvec{\Sigma}\of{\alpha}\notl\big)^{-1}
&= \big(\hvec{\Sigma}\of{\alpha}\train\big)^{-1} + 
   \frac{\big(\hvec{\Sigma}\of{\alpha}\train\big)^{-1} 
         \hvec{n}_l g_l\of{\alpha} \hvec{n}_l\herm 
         \big(\hvec{\Sigma}\of{\alpha}\train\big)^{-1}}
   {1-g_l\of{\alpha} \hvec{n}_l\herm \big(\hvec{\Sigma}\of{\alpha}\train\big)^{-1} \hvec{n}_l}
\label{eq:invSignotl}
\end{align}
for
\begin{align}
\hvec{\Sigma}\of{\alpha}\train
&\defn (1-\alpha)\frac{Q}{Q-1} \hvec{\Sigma}\train + \alpha c\vec{I}_M \\
g_l\of{\alpha}
&\defn (1-\alpha)\frac{1}{Q-1}\left(1+\frac{|s_l|^2}{\|\vec{s}\|^2-|s_l|^2}\right) .
\end{align}
Merging \eqref{wmf_alpha}, \eqref{hnotl}, and \eqref{invSignotl}, we find that
\begin{align}
\hat{r}_l\of{\alpha}
&= \vec{y}_l\herm\big(\hvec{\Sigma}\of{\alpha}\train\big)^{-1}\hvec{h}\train
  + \frac{\vec{y}_l\herm\big(\hvec{\Sigma}\train\of{\alpha}\big)^{-1}\hvec{n}_l}
         {1-g_l\of{\alpha}\hvec{n}_l\herm\big(\hvec{\Sigma}\train\of{\alpha}\big)^{-1}\hvec{n}_l} 
\nonumber\\&\quad\times
    \left(g_l\of{\alpha}\hvec{n}_l\herm \big(\hvec{\Sigma}\train\of{\alpha}\big)^{-1}\hvec{h}\train
          - \frac{s_l}{\|\vec{s}\train\|^2-|s_l|^2}\right) .
\end{align}
With the eigen-decomposition $\hvec{\Sigma}\train = \vec{V}\train\vec{\Lambda}\train\vec{V}\train\herm$, we have
\begin{align}
\big(\hvec{\Sigma}\of{\alpha}\train\big)^{-1}
&= \vec{V}\train\bigg( 
        \underbrace{ (1-\alpha)\frac{Q}{Q-1} \vec{\Lambda}\train + \alpha c\vec{I}_M }_{
                \displaystyle \defn \Diag(\vec{\gamma}\of{\alpha})}
        \bigg)^{-1} \vec{V}\train\herm  
\end{align}
which can be used to compute 
$\hvec{r}\train\of{\alpha}=[\hat{r}_1\of{\alpha},\dots,\hat{r}_Q\of{\alpha}]\tran$ 
efficiently via 
\begin{align}
\vec{y}_l\herm\big(\hvec{\Sigma}\of{\alpha}\train\big)^{-1}\hvec{h}\train
&= \Big[ \vec{Y}\train\herm \vec{V}\train 
        \big( \vec{V}\train\herm \hvec{h}\train \oslash \vec{\gamma}\of{\alpha}\big)
        \Big]_l \\
\hvec{n}_l\herm\big(\hvec{\Sigma}\of{\alpha}\train\big)^{-1}\hvec{h}\train
&= \Big[ \hvec{N}\train\herm \vec{V}\train 
        \big( \vec{V}\train\herm \hvec{h}\train \oslash \vec{\gamma}\of{\alpha}\big) 
        \Big]_l 
\end{align}
\begin{align}
&= \Big[ \Big( \big(\vec{Y}\train\herm\vec{V}\train\big)\odot\big(\hvec{N}\train\herm\vec{V}\train\big)^* \Big) 
   \big(\vec{1}\oslash\vec{\gamma}\of{\alpha}\big) \Big]_l,
\end{align}
where $\hvec{N}\train\defn\vec{Y}\train-\hvec{h}\train\vec{s}\train\herm$ is an estimate of the interference $\vec{N}\train$, $\odot$ denotes element-wise multiplication, and $\oslash$ denotes element-wise division.

For a given $\alpha$, the unbiasing gain $\beta\of{\alpha}$ (recall \eqref{wmf_unbiased}) obeys
\begin{align}
\E\big\{\beta\of{\alpha}\hat{r}_l\of{\alpha} \big| s_l\big\} 
&= s_l, ~~l\in\{1,\dots,Q\} ,
\end{align}
and thus can be estimated as
\begin{align}
\beta\of{\alpha} 
&= \frac{1}{\E\big\{\hat{r}_l\of{\alpha}/s_l\big\}}
\approx \frac{Q}{\sum_{l=1}^Q \hat{r}_l\of{\alpha}/s_l} \defn \hat{\beta}\of{\alpha}.
\end{align}
After scaling by $\hat{\beta}\of{\alpha}$, the error precision $\hat{\xi}\of{\alpha}$ is 
\begin{align}
\hat{\xi}\of{\alpha}
&= \frac{1}{\frac{1}{Q}\sum_{l=1}^Q \big|\hat{\beta}\of{\alpha}\hat{r}_l\of{\alpha} - s_l\big|^2} .
\end{align}
The value of $\alpha$ can be optimized by maximizing $\hat{\xi}\of{\alpha}$ over a grid of possible values.

\section{GLRT via Deterministic Interference} \label{sec:McWhorter}

We now propose a different adaptive detector for $\vec{s}\sim p(\vec{s})$ that treats the interference $\vec{B\Phi}\herm$ as a deterministic unknown, rather than as temporally white and Gaussian, as in \secref{GerlachSteiner}.
In particular, it treats $\vec{B}\in\Complex^{M\times N}$ and $\vec{\Phi}\in\Complex^{L\times N}$ as deterministic unknowns, as in \cite{McWhorter:ASAP:04}, for some rank hypothesis $N<\min\{M,L\}$.
The rank hypothesis $N$ will be adapted as described in \secref{rankMc}.
However, we first describe the approach under a fixed choice of $N$.
In this case, the binary hypothesis test \eqref{H1H0} implies the GLRT 
\begin{align}
\frac{\max_{\vec{h},\vec{B},\vec{\Phi},\nu>0} p(\vec{Y}|\mc{H}_1;\vec{h},\vec{B},\vec{\Phi},\nu)}
{\max_{\vec{B},\vec{\Phi},\nu>0} p(\vec{Y}|\mc{H}_0;\vec{B},\vec{\Phi},\nu)} 
\gtreqless \eta
\label{eq:glrtb}.
\end{align}

\subsection{GLRT Denominator} \label{sec:GLRTdenMc}

Starting with the denominator of \eqref{glrtb}, we have
\begin{align}
p(\vec{Y}|\mc{H}_0;\vec{B},\vec{\Phi},\nu)
&= \prod_{m=1}^M \frac{\exp\big(-\|\vec{y}_m\herm - \vec{b}_m\herm \vec{\Phi}\herm\|^2/\nu\big)}{(\pi \nu)^L} 
\label{eq:pY0b_1},
\end{align}
where $\vec{y}_m\herm$ denotes the $m$th row of $\vec{Y}$ and
$\vec{b}_m\herm$ denotes the $m$th row of $\vec{B}$.
Due to the factorization in \eqref{pY0b_1}, the ML estimate of each $\vec{b}_m$ can be individually computed as 
\begin{align}
\hvec{b}_{0,m}
&\defn \argmin_{\vec{b}_m} \|\vec{y}_m-\vec{\Phi b}_m\|^2
= \vec{\Phi}^+ \vec{y}_m ,
\end{align}
where $(\cdot)^+$ denotes the pseudo-inverse, i.e., 
$\vec{\Phi}^+=(\vec{\Phi}\herm\vec{\Phi})^{-1}\vec{\Phi}\herm$.
Plugging $\hvec{b}_{0,m}$ into \eqref{pY0b_1} gives
\begin{align}
p(\vec{Y}|\mc{H}_0;\hvec{B}_0,\vec{\Phi},\nu)
&= \prod_{m=1}^M \frac{\exp\big(-\|\vec{y}_m\herm \vec{P}_{\vec{\Phi}}^\perp\|^2/\nu\big)}{(\pi \nu)^L} \\
&= \frac{\exp\big(-\tr\{\vec{Y}\vec{P}_{\vec{\Phi}}^\perp\vec{Y}\herm\}/\nu\big)}{(\pi \nu)^{ML}} 
\label{eq:pY0b_2} .
\end{align}
Next we maximize over the noise variance $\nu>0$.
The negative log-likelihood is
\begin{align}
\lefteqn{ -\ln p(\vec{Y}|\mc{H}_0;\hvec{B}_0,\vec{\Phi},\nu) }\nonumber\\
&= \tr\{\vec{Y}\vec{P}_{\vec{\Phi}}^\perp\vec{Y}\herm\}/\nu + ML\ln\pi + ML\ln \nu
\label{eq:logpY0b_0},
\end{align}
and so zeroing its gradient gives the ML estimate
\begin{align}
\hat{\nu}_0
&= \frac{1}{ML} \tr\{\vec{Y}\vec{P}_{\vec{\Phi}}^\perp\vec{Y}\herm\} 
\label{eq:nu0}.
\end{align}
Plugging this back into \eqref{logpY0b_0} gives
\begin{align}
\lefteqn{ -\ln p(\vec{Y}|\mc{H}_0;\hvec{B}_0,\vec{\Phi},\hat{\nu}_0) }\nonumber\\
&= ML(1+\ln\pi) + ML\ln\left(\frac{1}{ML}\tr\{\vec{Y}\vec{P}_{\vec{\Phi}}^\perp\vec{Y}\herm\}\right)
\label{eq:logpY0b_1} .
\end{align}
Finally, minimizing this negative log-likelihood over $\vec{\Phi}$ is equivalent to minimizing $\tr\{\vec{Y}\vec{P}_{\vec{\Phi}}^\perp\vec{Y}\herm\} = \tr\{\vec{YY}\herm\} - \tr\{\vec{Y}\vec{P}_{\vec{\Phi}}\vec{Y}\herm\}$, or maximizing $\tr\{\vec{Y}\vec{P}_{\vec{\Phi}}\vec{Y}\herm\}=\tr\{\vec{P}_{\vec{\Phi}}\vec{Y}\herm\vec{Y}\vec{P}_{\vec{\Phi}}\}$.
But since the trace of a matrix is the sum of its eigenvalues, 
the optimal $\vec{\Phi}$ are those whose column space is the span of the dominant eigenvectors of $\vec{Y}\herm\vec{Y}$.
In summary, the minimized negative log-likelihood equals
\begin{align}
\lefteqn{ -\ln p(\vec{Y}|\mc{H}_0;\hvec{B}_0,\hvec{\Phi}_0,\hat{\nu}_0) }\nonumber\\
&= ML(1+\ln\pi) + ML\ln\left(\frac{1}{M} \sum_{m=N+1}^M \lambda_{0,m} \right) 
\label{eq:logpY0b} ,
\end{align}
where $\{\lambda_{0,m}\}_{m=1}^M$ are the eigenvalues of $\frac{1}{L}\vec{Y}\herm\vec{Y}$ in decreasing order, as per \eqref{lambda}.

\subsection{GLRT Numerator} \label{sec:GLRTnumMc}

For the numerator of \eqref{glrtb}, equation \eqref{H1} implies
\begin{align}
\lefteqn{
p(\vec{Y}|\mc{H}_1;\vec{h},\vec{B},\vec{\Phi},\nu) 
}\nonumber\\
&= \int p(\vec{Y}|\vec{s},\mc{H}_1;\vec{h},\vec{B},\vec{\Phi},\nu) \,p(\vec{s}) \dif\vec{s} \\
&= \int 
\frac{\exp(-\|\vec{Y}-\vec{B\Phi}\herm -\vec{hs}\herm\|_F^2/\nu)}{(\pi \nu)^{ML}}
p(\vec{s}) \dif\vec{s} 
\label{eq:pY1b}  
\end{align}
Exact maximization of $p(\vec{Y}|\mc{H}_1;\vec{h},\vec{B},\vec{\Phi},\nu)$ over 
\begin{align}
\vec{\Theta} &\defn \{\vec{h},\vec{B},\vec{\Phi},\nu\}
\label{eq:Theta}
\end{align}
appears to be intractable.
As before, we propose to apply EM with hidden data $\vec{s}$, which implies iterating 
\begin{align}
\hvec{\Theta}\of{t+1}
&= \arg\max_{\vec{\Theta}} \E\big\{\ln p(\vec{Y},\vec{s}|\mc{H}_1;\vec{\Theta}) \,\big|\, \vec{Y}; \hvec{\Theta}\of{t}\big\}
\label{eq:EMb} .
\end{align}

Because \textb{$\vec{s}$ is statistically independent of $\vec{\Theta}$}, \eqref{EMb} can be rewritten as 
\begin{align}
\hvec{\Theta}\of{t+1}
&= \arg\min_{\vec{\Theta}} \int \bigg[
\frac{\|\vec{Y}-\vec{B\Phi}\herm -\vec{hs}\herm\|_F^2}{\nu} 
+ ML \ln (\pi \nu)
\bigg] 
\nonumber\\&\hspace{22mm}\times
p(\vec{s}|\vec{Y};\hvec{\Theta}\of{t}) \dif\vec{s} \\
&= \arg\min_{\vec{\Theta}} \int 
\frac{\|\vec{Y}-\vec{B\Phi}\herm -\vec{hs}\herm\|_F^2}{\nu} 
p(\vec{s}|\vec{Y};\hvec{\Theta}\of{t}) \dif\vec{s} 
\nonumber\\&\hspace{16mm}
+ ML \ln (\pi \nu) 
\label{eq:Mstep1b}.
\end{align}
Noting that 
\begin{align}
\lefteqn{
\|\vec{Y}-\vec{B\Phi}\herm -\vec{hs}\herm\|_F^2
}\nonumber\\
&= \|\vec{Y}-\vec{B\Phi}\herm\|_F^2
   + \|\vec{h}\|^2 \|\vec{s}\|^2
\nonumber\\&\quad
   - \vec{h}\herm(\vec{Y}-\vec{B\Phi}\herm)\vec{s}
   - \vec{s}\herm(\vec{Y}\herm-\vec{\Phi}\vec{B}\herm)\vec{h} ,
\end{align}
we can rewrite \eqref{Mstep1b} as
\begin{align}
\hvec{\Theta}\of{t+1}
&= \arg\min_{\vec{\Theta}} \Big\{ 
   \frac{1}{\nu}\Big[ \|\vec{Y}-\vec{B\Phi}\herm\|_F^2 
   + \|\vec{h}\|^2 E\of{t}
\nonumber\\&\quad
   - \vec{h}\herm(\vec{Y}-\vec{B\Phi}\herm)\hvec{s}\of{t}
   - \hvec{s}\ofH{t}(\vec{Y}\herm-\vec{\Phi}\vec{B}\herm)\vec{h} \Big]
\nonumber\\&\quad
   + ML \ln (\pi \nu) 
   \Big\}
\label{eq:Mstep2b}
\end{align}
where, similar to before,
\begin{align}
\hvec{s}\of{t}
&\defn \E\{\vec{s}|\vec{Y};\hvec{\Theta}\of{t}\} 
\label{eq:shatb}\\
E\of{t} 
&\defn \E\{\|\vec{s}\|^2|\vec{Y};\hvec{\Theta}\of{t}\} 
\label{eq:Eb}.
\end{align}

We are now ready to minimize \eqref{Mstep2b} over $\vec{\Theta}=\{\vec{h},\vec{B},\vec{\Phi},\nu\}$.
Zeroing the gradient of the cost over $\vec{h}$ yields
\begin{align}
\hvec{h}\of{t+1}
&= \frac{ (\vec{Y}-\vec{B\Phi}\herm)\hvec{s}\of{t} }{ E\of{t} } 
\label{eq:hhatb}.
\end{align}
Plugging this back into \eqref{Mstep2b}, the term relevant to the optimization of $\vec{B}$ and $\vec{\Phi}$ becomes
\begin{align}
\lefteqn{
\|\vec{Y}-\vec{B\Phi}\herm\|_F^2 
 - \|(\vec{Y}-\vec{B\Phi}\herm)\hvec{s}\of{t}\|^2 / E\of{t} 
}\nonumber\\
&= \tr\big\{ (\vec{Y}-\vec{B\Phi}\herm)\tvec{P}^\perp_{\hvec{s}\of{t}}(\vec{Y}-\vec{B\Phi}\herm)\herm \big\}
\label{eq:Mstep3b},
\end{align}
with $\tvec{P}^\perp_{\hvec{s}\of{t}}$ from \eqref{Ptilsperp}.
To optimize \eqref{Mstep3b} over $\vec{B}$, we expand 
\begin{align}
\lefteqn{
\tr\big\{ (\vec{Y}-\vec{B\Phi}\herm)\tvec{P}^\perp_{\hvec{s}\of{t}}(\vec{Y}-\vec{B\Phi}\herm)\herm \big\}
}\nonumber\\
&= \text{const}
-\tr\big\{ \vec{B\Phi}\herm\tvec{P}^\perp_{\hvec{s}\of{t}}\vec{Y}\herm \big\} 
\nonumber\\&\quad
-\tr\big\{ \vec{Y}\tvec{P}^\perp_{\hvec{s}\of{t}}\vec{\Phi}\vec{B}\herm \big\}
+\tr\big\{ \vec{B\Phi}\herm\tvec{P}^\perp_{\hvec{s}\of{t}}\vec{\Phi}\vec{B}\herm\big\} ,
\end{align}
evaluate its gradient, which equals
\begin{align}
-2 \vec{Y}\tvec{P}^\perp_{\hvec{s}\of{t}}\vec{\Phi}
+2 \vec{B\Phi}\herm\tvec{P}^\perp_{\hvec{s}\of{t}}\vec{\Phi} ,
\end{align}
and set it to zero, yielding
\begin{align}
\hvec{B}\of{t+1}
&= \vec{Y}\tvec{P}^\perp_{\hvec{s}\of{t}}\vec{\Phi}
\big( \vec{\Phi}\herm\tvec{P}^\perp_{\hvec{s}\of{t}}\vec{\Phi} \big)^{-1} .
\end{align}
Plugging this back into \eqref{Mstep3b} gives
\begin{align}
\lefteqn{
\tr\big\{ (\vec{Y}-\hvec{B}\of{t+1}\vec{\Phi}\herm)
\tvec{P}^\perp_{\hvec{s}\of{t}}
(\vec{Y}-\hvec{B}\of{t+1}\vec{\Phi}\herm)\herm \big\} 
}\nonumber\\
&= 
\tr\big\{ 
\vec{Y}
(\vec{I}-\tvec{P}^\perp_{\hvec{s}\of{t}}\vec{\Phi}
         [\vec{\Phi}\herm\tvec{P}^\perp_{\hvec{s}\of{t}}\vec{\Phi}]^{-1}
         \vec{\Phi}\herm)
\tvec{P}^\perp_{\hvec{s}\of{t}}
\nonumber\\&\quad\qquad\times
(\vec{I}-\tvec{P}^\perp_{\hvec{s}\of{t}}\vec{\Phi}
         [\vec{\Phi}\herm\tvec{P}^\perp_{\hvec{s}\of{t}}\vec{\Phi}]^{-1}
         \vec{\Phi}\herm)
\vec{Y}\herm
\big\} \\
&= 
\tr\big\{ 
\ovec{Y}
(\vec{I}-\ovec{\Phi}
         [\ovec{\Phi}\herm\ovec{\Phi}]^{-1}
         \ovec{\Phi}\herm)^2
\ovec{Y}\herm
\big\} \\
&= 
\tr\big\{ 
\ovec{Y}
\vec{P}^\perp_{\ovec{\Phi}}
\ovec{Y}\herm
\big\} 
\label{eq:Mstep4b} ,
\end{align}
with $\ovec{Y}\defn \vec{Y}(\tvec{P}^\perp_{\hvec{s}\of{t}})^{\frac{1}{2}}$ 
and $\ovec{\Phi}\defn (\tvec{P}^\perp_{\hvec{s}\of{t}})^{\frac{1}{2}}\vec{\Phi}$.
From \eqref{Ptilsperp}, note
\begin{align}
(\tvec{P}^\perp_{\hvec{s}\of{t}})^{\frac{1}{2}}
&= \vec{I} + \big(\zeta\of{t}-1\big) \vec{P}_{\hvec{s}\of{t}} 
\label{eq:Ptilsperp_sqrt}\\
\zeta\of{t}
&\defn \sqrt{ 1-\|\hvec{s}\of{t}\|^2/E\of{t} }
\label{eq:zeta} .
\end{align}
The $\ovec{\Phi}$ that minimize $\tr\{\ovec{Y}\vec{P}^\perp_{\ovec{\Phi}}\ovec{Y}\herm\}=\tr\{\vec{P}^\perp_{\ovec{\Phi}}\ovec{Y}\herm\ovec{Y}\vec{P}^\perp_{\ovec{\Phi}}\}$ are those whose column space equals the span of the $N$ dominant eigenvectors of $\ovec{Y}\herm\ovec{Y}$, and so
\begin{align}
\min_{\ovec{\Phi}} \tr\{\ovec{Y}\vec{P}^\perp_{\ovec{\Phi}}\ovec{Y}\herm\}
&= \sum_{m=N+1}^M \lambda_m(\ovec{Y}\herm\ovec{Y})
\end{align}
where $\lambda_m(\ovec{Y}\herm\ovec{Y})$ is the $m$th eigenvalue of $\ovec{Y}\herm\ovec{Y}$ in decreasing order.
These eigenvalues are the same as those of 
\begin{align}
\ovec{YY}\herm 
&= \vec{Y}\tvec{P}_{\hvec{s}\of{t}}^\perp\vec{Y}\herm .
\end{align}
Thus, the optimization \eqref{Mstep2b} reduces to 
\begin{align}
\hat{\nu}_1\of{t+1}
&= \arg\min_\nu \Big\{ 
   \frac{L}{\nu}
   \sum_{m=N+1}^M \lambda_{1,m}\of{t+1}
   + ML \ln (\pi \nu) 
   \Big\}
\label{eq:Mstep5b},
\end{align}
where $\{\lambda_{1,m}\of{t+1}\}_{m=1}^M$ are the eigenvalues of $\tfrac{1}{L}\vec{Y}\tvec{P}_{\hvec{s}\of{t}}^\perp\vec{Y}\herm$ in decreasing order.
Zeroing the derivative of \eqref{Mstep5b} w.r.t.\ $\nu$ yields
\begin{align}
\hat{\nu}_1\of{t+1}
&= \frac{1}{M} \sum_{m=N+1}^M \lambda_{1,m}\of{t+1} .
\end{align}
Plugging $\hat{\nu}_1\of{t+1}$ back into the cost expression yields the iteration-$(t\!+\!1)$ EM-maximized log-likelihood under $\mc{H}_1$:
\begin{align}
\lefteqn{
\ln p(\vec{Y}|\mc{H}_1;\hvec{\Theta}\of{t+1}) 
}\nonumber\\
&= ML (1+\ln \pi) + ML \ln\left( \frac{1}{M} \sum_{m=N+1}^M \lambda_{1,m}\of{t+1} \right) .
\end{align}

\subsection{EM Update under an Independent Prior}

The EM updates \eqref{shatb}-\eqref{Eb} depend on the choice of $p(\vec{s})$.
For an independent prior, as in \eqref{s_indep}, we can compute the MMSE estimate of the $l$th symbol using the measurement equation
\begin{align}
\vec{y}_l
&= \hvec{h}\of{t} s_l^* + \hvec{B}\of{t}\hvec{\phi}_l\of{t} + \mc{CN}(\vec{0},\hat{\nu}_1\of{t}\vec{I}) ,
\end{align}
where $\hvec{\phi}_l$ denotes the $l$th column of $\hvec{\Phi}\herm$. 
From $\vec{y}_l$, we can obtain the following sufficient statistic \cite{Scharf:Book:91} for the estimation of $s_l$ through matched filtering, i.e.,
\begin{align}
\lefteqn{
\tilde{r}_l\of{t}
\defn \hvec{h}\ofH{t}\vec{y}_l
}\\
&= \|\hvec{h}\of{t}\|^2 s_l^* 
   + \hvec{h}\ofH{t}\hvec{B}\of{t}\hvec{\phi}_l\of{t} 
   + \mc{CN}(0,\hat{\nu}_1\of{t}\|\hvec{h}\of{t}\|^2 ) .
\end{align}
We find it more convenient to work with the shifted, conjugated, and normalized statistic
\begin{align}
r_l\of{t}
&\defn \frac{1}{\|\hvec{h}\of{t}\|^2} 
       \big(\tilde{r}_l\of{t} 
            - \hvec{h}\of{t}\hvec{B}\of{t}\hvec{\phi}_l\of{t}\big)^* \\
&= s_l + \mc{CN}\bigg(0,\frac{\hat{\nu}_1\of{t}}{\|\hvec{h}\of{t}\|^2} \bigg) ,
\end{align}
noting that 
\begin{align}
\vec{r}\ofH{t}
&= \frac{1}{\|\hvec{h}\of{t}\|^2} \hvec{h}\ofH{t}
   \big( \vec{Y}-\hvec{B}\of{t}\hvec{\Phi}\ofH{t} \big)
\label{eq:r}.
\end{align}

To efficiently compute \eqref{r}, we first note that
\begin{align}
\lefteqn{
\vec{Y} -\hvec{B}\of{t}\hvec{\Phi}\ofH{t}
}\nonumber\\
&= \vec{Y} - \vec{Y}
        \tvec{P}^\perp_{\hvec{s}}\hvec{\Phi}
        \big( \hvec{\Phi}\herm
              \tvec{P}^\perp_{\hvec{s}}
              \hvec{\Phi}
              \big)^{-1} 
   \hvec{\Phi}\herm \\
&= \vec{Y}\big[ \vec{I} -
        \tvec{P}^\perp_{\hvec{s}}\hvec{\Phi}
        \big( \hvec{\Phi}\herm
              \tvec{P}^\perp_{\hvec{s}}
              \hvec{\Phi}
              \big)^{-1} 
   \hvec{\Phi}\herm
   \big] 
   \end{align}  
   \begin{align} 
&= \vec{Y} (\tvec{P}^\perp_{\hvec{s}})^{\frac{1}{2}}
   \big[ \vec{I} -
        (\tvec{P}^\perp_{\hvec{s}})^{\frac{1}{2}}
        \hvec{\Phi}
        \big( \hvec{\Phi}\herm
              \tvec{P}^\perp_{\hvec{s}}
              \hvec{\Phi}
              \big)^{-1} 
   \hvec{\Phi}\herm (\tvec{P}^\perp_{\hvec{s}})^{\frac{1}{2}}
   \big] 
   (\tvec{P}^\perp_{\hvec{s}})^{-\frac{1}{2}} \\
&= \ovec{Y}\big[ \vec{I} - 
   \ovec{\Phi}\big(\ovec{\Phi}\herm\ovec{\Phi}\big)^{-1}\ovec{\Phi}\herm
   \big]
   (\tvec{P}^\perp_{\hvec{s}})^{-\frac{1}{2}} \\
&= \ovec{Y} \vec{P}_{\ovec{\Phi}}^\perp
   (\tvec{P}^\perp_{\hvec{s}})^{-\frac{1}{2}} 
\label{eq:YBPhi} ,
\end{align}
where we omitted the time index for brevity
and defined
\begin{align}
\ovec{\Phi}\of{t}
&\defn ( \tvec{P}^\perp_{\hvec{s}\of{t-1}})^{\frac{1}{2}} \hvec{\Phi}\of{t} \\
\ovec{Y}\of{t}
&\defn \vec{Y}(\tvec{P}^\perp_{\hvec{s}\of{t-1}})^{\frac{1}{2}} 
\label{eq:Ybar} ,
\end{align}
noting that \eqref{Ptilsperp_sqrt}-\eqref{zeta} imply 
\begin{align}
(\tvec{P}^\perp_{\hvec{s}\of{t}})^{-\frac{1}{2}}
&= \vec{I}_L + \Big(\frac{1}{\zeta\of{t}}-1\Big) \vec{P}_{\hvec{s}\of{t}} 
\label{eq:Ptilsperp_invsqrt} .
\end{align}
Suppose we take the singular value decomposition (SVD)
\begin{align}
\ovec{Y}\of{t}
&=\vec{V}\of{t}\vec{D}_1\of{t}\vec{U}\ofH{t} ,
\end{align}
where 
\begin{align}
\diag\big(\vec{D}_1\of{t}\big)
=\Big[\sqrt{L\lambda_{1,1}\of{t}},\dots,\sqrt{L\lambda_{1,M}\of{t}}\Big]\tran
\end{align}
with $\lambda_{1,m}\of{t}$ defined after \eqref{Mstep5b}.
Then, using the fact that
the column space of $\ovec{\Phi}\of{t}$ spans the $N$-dimensional principal eigenspace of $\ovec{Y}\ofH{t} \ovec{Y}\of{t}$
(as discussed after \eqref{Mstep4b}),
we have 
\begin{align}
\ovec{Y}\of{t}\vec{P}_{\ovec{\Phi}\of{t}}^\perp
&=\ovec{Y}\of{t}(\vec{I}-\vec{P}_{\ovec{\Phi}\of{t}}) \\
&=\vec{Y}(\tvec{P}^\perp_{\hvec{s}\of{t-1}})^{\frac{1}{2}} 
- \ovec{V}\of{t}\ovec{D}_1\of{t}\ovec{U}\ofH{t} ,
\end{align}
where 
$\ovec{V}\of{t}\in\Complex^{M\times N}$,
$\ovec{D}_1\of{t}\in\Real^{N\times N}$, and
$\ovec{U}\of{t}\in\Complex^{L\times N}$ 
contain the $N$ principal components of 
$\vec{V}\of{t}$, $\vec{D}_1\of{t}$, and $\vec{U}\of{t}$. 
Plugging this into \eqref{YBPhi}, we get
\begin{align}
\lefteqn{
\vec{Y}-\hvec{B}\of{t}\hvec{\Phi}\ofH{t}
}\nonumber\\
&=\vec{Y}
- \ovec{V}\of{t}\ovec{D}_1\of{t}\ovec{U}\ofH{t} 
(\tvec{P}^\perp_{\hvec{s}\of{t-1}})^{-\frac{1}{2}} 
\label{eq:YbarPperp} \\
&=\Big[ \ovec{Y}\of{t}
- \ovec{V}\of{t}\ovec{D}_1\of{t}\ovec{U}\ofH{t} \Big]
(\tvec{P}^\perp_{\hvec{s}\of{t-1}})^{-\frac{1}{2}} 
\label{eq:YbarPperp2} .
\end{align}

Applying \eqref{YbarPperp} to \eqref{hhatb} yields
\begin{eqnarray}
\lefteqn{
\hvec{h}\of{t}
= \frac{1}{E\of{t-1}} (\vec{Y}-\hvec{B}\of{t}\hvec{\Phi}\ofH{t})\hvec{s}\of{t-1}
}\\
&=& \frac{1}{E\of{t-1}} \Big(\vec{Y}\hvec{s}\of{t-1} - \frac{1}{\zeta\of{t-1}} 
   \ovec{V}\of{t}\ovec{D}_1\of{t}\ovec{U}\ofH{t} 
   \hvec{s}\of{t-1} 
   \Big) 
\label{eq:hhat2b}, \qquad
\end{eqnarray}
and applying \eqref{YbarPperp2} to \eqref{r} yields
\begin{align}
\vec{r}\ofH{t}
&= \frac{1}{\|\hvec{h}\of{t}\|^2} 
   \hvec{h}\ofH{t} 
   \Big[ \ovec{Y}\of{t} 
        - \ovec{V}\of{t}\ovec{D}_1\of{t}\ovec{U}\ofH{t} \Big]
   (\tvec{P}^\perp_{\hvec{s}\of{t-1}})^{-\frac{1}{2}} 
\label{eq:r1} .
\end{align}
We can simplify the previous expression by noting that
\begin{align}
\lefteqn{
\Big[ \ovec{Y}\of{t} 
- \ovec{V}\of{t}\ovec{D}_1\of{t}\ovec{U}\ofH{t} 
\Big] (\tvec{P}^\perp_{\hvec{s}\of{t-1}})^{-\frac{1}{2}}
}\nonumber\\
&=
\Big[ \ovec{Y}\of{t}
- \ovec{V}\of{t}\ovec{D}_1\of{t}\ovec{U}\ofH{t} 
\Big] 
\Big(
\vec{I}_L + \frac{1-\zeta\of{t-1}}{\zeta\of{t-1}} \vec{P}_{\hvec{s}\of{t-1}}
\Big) 
\label{eq:shat2b} 
   \end{align} 
   \begin{align} 
&=
\ovec{Y}\of{t} - \ovec{V}\of{t}\ovec{D}_1\of{t}\ovec{U}\ofH{t} 
+ 
\frac{1-\zeta\of{t-1}}{\zeta\of{t-1}\|\hvec{s}\of{t-1}\|^2} 
\nonumber\\&\quad\times
\Big[ \ovec{Y}\of{t}\hvec{s}\of{t-1} 
- \ovec{V}\of{t}\ovec{D}_1\of{t}\ovec{U}\ofH{t} \hvec{s}\of{t-1}
\Big] \hvec{s}\ofH{t-1} \\
&=
\ovec{Y}\of{t} - \ovec{V}\of{t}\ovec{D}_1\of{t}\ovec{U}\ofH{t} 
+ 
\frac{1-\zeta\of{t-1}}{\|\hvec{s}\of{t-1}\|^2} 
\label{eq:shat3b} \\&\quad\times
\Big[ \vec{Y}\hvec{s}\of{t-1} 
- \frac{1}{\zeta\of{t-1}} \ovec{V}\of{t}\ovec{D}_1\of{t}\ovec{U}\ofH{t} \hvec{s}\of{t-1}
\Big] \hvec{s}\ofH{t-1} 
\nonumber \\
&=
\ovec{Y}\of{t} - \ovec{V}\of{t}\ovec{D}_1\of{t}\ovec{U}\ofH{t} 
+ \frac{(1-\zeta\of{t-1})E\of{t-1}}{\|\hvec{s}\of{t-1}\|^2} 
\hvec{h}\of{t}\hvec{s}\ofH{t-1} 
\label{eq:shat4b} \\
&=
\ovec{Y}\of{t} - \ovec{V}\of{t}\ovec{D}_1\of{t}\ovec{U}\ofH{t} 
+ \frac{1}{1+\zeta\of{t-1}} \hvec{h}\of{t}\hvec{s}\ofH{t-1} 
\label{eq:shat5b},
\end{align}
where \eqref{shat2b} used \eqref{Ptilsperp_invsqrt};
\eqref{shat3b} used the fact that $\ovec{Y}\of{t}\hvec{s}\of{t-1}=\zeta\of{t-1}\vec{Y}\hvec{s}\of{t-1}$, as implied by \eqref{Ybar} and \eqref{Ptilsperp_sqrt};
\eqref{shat4b} used \eqref{hhat2b}; and 
\eqref{shat5b} used \eqref{zeta}.
Plugging \eqref{shat5b} into \eqref{r1} then yields
\begin{align}
\hvec{r}\ofH{t}
&= \frac{1}{\|\hvec{h}\of{t}\|^2} \hvec{h}\ofH{t} 
\Big( \ovec{Y}\of{t} 
- \ovec{V}\of{t}\ovec{D}_1\of{t}\ovec{U}\ofH{t} 
\Big)
\nonumber\\&\quad
+ \frac{1}{1+\zeta\of{t-1}} \hvec{s}\ofH{t-1} . 
\end{align}

Given $\vec{r}\of{t}$, the computation of $\hvec{s}\of{t}$ and $E\of{t}$ follows the procedure discussed around \eqref{shatl_gauss}-\eqref{E_discrete}.
This EM update procedure is summarized in \algref{McW}.

\begin{algorithm}[t]
\footnotesize
\caption{EM update under deterministic interference}
\label{alg:McW}
\begin{algorithmic}[1]
\REQUIRE{
    Data $\vec{Y}\in\Complex^{M\times L}$,
    signal prior $p(\vec{s})=\prod_{l=1}^L p_l(s_l)$.
}
\smallskip
\STATE
Initialize $\hvec{s}$ and $E>0$ (see \secref{initGS})
\smallskip
\REPEAT
\smallskip
  \STATE 
    $\zeta \leftarrow \sqrt{1-\|\hvec{s}\|^2/E}$
  \STATE
    $\vec{g} \leftarrow \vec{Y}\hvec{s}/\|\hvec{s}\|^2$
  \STATE 
    $\ovec{Y} \leftarrow \vec{Y} + (\zeta-1)\vec{g}\hvec{s}\herm$
  \STATE
    Estimate interference rank $N$ (see \secref{rankMc}).
  \STATE 
    $\big\{\ovec{V},\ovec{D}_1,\ovec{U}\herm\big\}
     \leftarrow \text{principal\_svd}\big(\ovec{Y}
                ,N\big)$
  \STATE 
    $\hat{\nu}_1 \leftarrow \frac{1}{ML}\Big(\|\ovec{Y}\|_F^2 
        - \tr\big\{\ovec{D}_1^2\big\}\Big)$
  \STATE 
    $\hvec{h} \leftarrow \frac{1}{E}\Big( \|\hvec{s}\|^2\vec{g} - 
        \frac{1}{\zeta}\ovec{V}\,\ovec{D}_1\,\ovec{U}\herm\hvec{s} \Big)$
  \STATE 
    $\xi \leftarrow \frac{\|\hvec{h}\|^2}{\hat{\nu}_1}$
  \STATE 
    $\vec{r}
     \leftarrow 
        \frac{1}{\|\hvec{h}\|^2}
        \big( \ovec{Y}\herm\hvec{h} 
        - \ovec{U}\,\ovec{D}_1 \ovec{V}\herm \hvec{h} \big)
        + \frac{1}{1+\zeta} 
          \hvec{s} 
        \text{,~where~} \vec{r} \sim \mc{CN}(\vec{s},\frac{1}{\xi}\vec{I})
     $
  \STATE 
    $\hat{s}_l \leftarrow \E\{s_l | r_l; \xi\}~\forall l=1,\dots,L$
  \STATE 
    $E \leftarrow \sum_{l=1}^L \E\{|s_l|^2 | r_l; \xi\}$
\smallskip
\UNTIL{Terminated}
\end{algorithmic}
\end{algorithm}

\subsection{Evaluation of the GLRT}

Denoting the final EM estimates by $\hvec{s}$ and $\hvec{\Theta}$, the (EM approximate) GLRT statistic, in the log domain, becomes 
\begin{align}
\ln p(\vec{Y}|\mc{H}_1;\hvec{\Theta}) 
-\ln p(\vec{Y}|\mc{H}_0;\hvec{\Theta}) 
= ML \ln \frac{\hat{\nu}_0}{\hat{\nu}_1}
\label{eq:glrt2b} ,
\end{align}
with $\nu_0$ computed from \eqref{nu0} and $\nu_1$ computed from \algref{McW}.

\textb{Although the proposed GLRT is not CFAR, neither is the simpler McWhorter \cite{McWhorter:ASAP:04} detector that results in the special case where $\vec{s}$ is known (i.e., $p(\vec{s})$ is a point mass).
So, to set the detection threshold $\eta$ (recall \eqref{glrtb}) in practice, one could run experiments or simulations to provide histograms of the test statistic under $\mc{H}_0$ and $\mc{H}_1$, and then choose the value of $\eta$ that yields the desired balance between miss rate and false-alarm rate.
}

\subsection{Estimating the Interference Rank} \label{sec:rankMc}

To estimate the interference rank $N=\rank(\vec{R})$, we adopt the same approach as described in \secref{rankGS}.
But now the DoF $D(N)$ of the parameters $\vec{\Theta}_N$ is different.
In particular, the DoF in $\vec{h}$ equals $2M$;
the DoF in $\vec{B}\vec{\Phi}\herm$, an $M\times L$ rank-$N$ complex-valued matrix, equals $2(M+L-N)N$;
and the DoF in the noise variance $\nu$ equals $1$.
In summary, $D(N)=2(M+L-N)N + 2M + 1$.
For our numerical experiments, we used GIC with $G=1.7$. 

%

\section{Numerical Experiments} \label{sec:num}

\blue
We now present numerical experiments to evaluate the proposed detectors.
The experiments focus on the signal-detection application in communications, 
as introduced in \secref{problem} and described in more detail below.

\subsection{Signal detection in communications} \label{sec:comms}

Consider the problem of detecting the presence or absence of a communications signal from $M$ antennas in the presence of $N$ interferers and white Gaussian noise.
Under the narrowband and slow-fading assumptions, the baseband received waveform at the $m$th antenna and time $t$ takes the form 
\cite{Meyr:Book:98} 
\begin{subequations} \label{eq:ymt}
\begin{align}
\mc{H}_1: y_m(t)
&= \tilde{h}_m e^{j (2\pi f_o t + \theta_o)} s^*(t-\tau_oT) 
\nonumber\\&\quad + \sum_{n=1}^N b_{mn} \phi_n(t) + w_m(t) 
\label{eq:ymtH1}\\
\mc{H}_0: y_m(t)
&= \sum_{n=1}^N b_{mn} \phi_n(t) + w_m(t) 
\label{eq:ymtH0}
\end{align}
\end{subequations}
under the signal-present (i.e., $\mc{H}_1$) and signal-absent (i.e., $\mc{H}_0$) hypothesis, respectively.
Here, 
$s(t)\in\Complex$ is the signal waveform,
$\phi_n(t)\in\Complex$ is the $n$th interference waveform, 
$\tilde{h}_m\in\Complex$ and $b_{mn}\in\Complex$ are baseband-equivalent channel gains,
and $w_m(t)$ is the noise waveform. 
Furthermore,
$f_o$ is the frequency offset (in Hz),
$\theta_o$ is the phase offset (in radians), 
and $\tau_o$ is the baud-normalized timing offset.
Under the standard assumption that the transmitter and receiver both use square-root raised-cosine pulse-shaping, we have \cite{Meyr:Book:98}
\begin{align}
s(t) 
&= \sum_{l=1}^L s_l g(t-lT) 
\label{eq:st}\\
g(t)
&= \frac{\cos(\alpha\pi t/T)}{1-(2\alpha t/T)} \frac{\sin(\pi t/T)}{\pi t/T} 
\label{eq:rc}.
\end{align}
where $s_l$ is a symbol from alphabet $\mc{A}_l\subset\Complex$,
$T$ is the baud interval, 
and
$g(t)\in\Real$ is a raised cosine (RC) pulse with roll-off factor $\alpha\in[0,1]$.

Suppose that we sample $y_m(t)$ every $T$ seconds, starting at time $t=\tau T$, where $\tau$ is a baud-normalized delay that we discuss in the sequel.
Under $\mc{H}_1$, this gives a matrix $\vec{Y}\of{\tau}\in\Complex^{M\times L}$
of space-time samples with entries
\begin{align}
[\vec{Y}\of{\tau}]_{ml} 
&= y_m(\tau T + lT) 
\\
&= \tilde{h}_m e^{j (2\pi f_o T (\tau+l) + \theta_o)} 
        s^*\big(\tau T + (l-\tau_o)T\big)
\nonumber\\&\quad 
   + \sum_{n=1}^N b_{mn} \phi_n\big(\tau T + lT\big) + w_m\big(\tau T + lT\big) \\
&= h_m\of{\tau} 
   \bigg[ \sum_{l'=1}^L  s_{l'}^* g\big((\tau+l-\tau_o-l')T\big) \bigg]
   e^{j 2\pi f_o T l} 
\nonumber\\&\quad + \sum_{n=1}^N b_{mn} \phi\of{\tau}_{ln} + w\of{\tau}_{ml} ,
\end{align}
where
$\phi\of{\tau}_{ln} \defn \phi_n(\tau T + lT)$,
$w\of{\tau}_{ml} \defn w_m(\tau T + lT)$, and
\begin{align}
h\of{\tau}_m &\defn \tilde{h}_m e^{j (2\pi f_o T \tau + \theta_o)} .
\end{align}
Thus with 
$\vec{h}\of{\tau}\defn[h_1\of{\tau},\dots,h_M\of{\tau}]\tran$,
$\vec{s}\defn[s_1,\dots,s_L]\tran$,
$\vec{B}\defn[b_{mn}]$, 
$\vec{\Phi}\of{\tau}\defn[\phi_{ln}\of{\tau}]$, and
$\vec{W}\of{\tau}\defn[w_{ml}\of{\tau}]$, 
we can write
\begin{align}
\vec{Y}\of{\tau}
&= \vec{h}\of{\tau} \vec{s}\herm \vec{G}_{\tau_o-\tau} \vec{J}_{f_o T} + \vec{B\Phi}\ofH{\tau} + \vec{W}\of{\tau}
\label{eq:Ytau} ,
\end{align}
where $\vec{J}_{f_o T}\in\Complex^{L\times L}$ is diagonal with 
$[\vec{J}_{\omega}]_{ll} \defn e^{j 2\pi \omega l}$
and $\vec{G}_{\tau_o-\tau}\in\Complex^{L\times L}$ is defined elementwise as
\begin{align}
[\vec{G}_{\Delta}]_{ql}
&\defn g((l-q-\Delta)T) .
\end{align}
Due to the square-root raised-cosine receiver filtering, 
each row of $\vec{W}\of{\tau}$ contains uncorrelated Gaussian noise samples \cite{Meyr:Book:98} for any $\tau$.
Thus, assuming that the noise is uncorrelated across antennas, the entries of $\vec{W}\of{\tau}$ are i.i.d.\ Gaussian.

Since $\vec{G}_0=\vec{I}$ due to the properties of the RC pulse \cite{Meyr:Book:98}, and since $\vec{J}_0=\vec{I}$ by inspection,
the space-time matrix $\vec{Y}\of{\tau}$ in \eqref{Ytau} matches $\vec{Y}$ in \eqref{H1} under perfect time synchronization (i.e., $\tau=\tau_o$) and perfect frequency synchronization (i.e., $f_o=0$).\footnote{The phase offset $\theta_o$ was absorbed into $\vec{h}\of{\tau}$, which we treat as a deterministic unknown during detection.}
But, in practice, oscillator mismatch ensures $f_o\neq 0$, and the unknown nature of $\tau_o$ ensures that $\tau\neq \tau_o$.

To alleviate the effects of time synchronization, we adopt the approach from \cite{Bliss:TSP:10}, which is to repeat the signal-detection test at many different delay hypotheses $\tau$.
In particular, we use the grid of delay hypotheses $\tau = k/P$, where $P$ is a fixed integer ``oversampling factor'' such as $P=2$, and where $k\in\Int_+$.
\emph{Thus, at each delay hypothesis $\tau=k/P$, we test for the presence or absence of a signal with true delay $\tau_o\approx k/P$.}
At the delay hypothesis $\tau=k/P$ closest to $\tau_o$ (i.e., $\tau=k_o/P$ with $k_o=\lfloor \tau_o P + \frac{1}{2}\rfloor$), the residual timing error is 
\begin{align}
\tilde{\tau_o} \defn \tau_o-\tfrac{k_o}{P}\in[-\tfrac{1}{2P},\tfrac{1}{2P}).
\end{align}
In this case, the space-time samples take the form
\begin{subequations} \label{eq:Ynum}
\begin{align}
\mc{H}_1: \vec{Y}
&= \vec{h} \vec{s}\herm \vec{G}_{\tilde{\tau}_o} \vec{J}_{f_o T} + \vec{B\Phi}\herm + \vec{W}
\label{eq:YnumH1} \\
\mc{H}_0: \vec{Y}
&= \vec{B\Phi}\herm + \vec{W}
\label{eq:YnumH0} 
\end{align}
\end{subequations}
with $\tilde{\tau}_o \in[-\tfrac{1}{2P},\tfrac{1}{2P})$.

\subsection{Experimental setup} \label{sec:setup}

For the numerical experiments in the sequel, we used \eqref{Ynum} with 
$\tilde{\tau}_o\sim\mc{U}[-\tfrac{1}{2P},\tfrac{1}{2P})$ and\footnote{%
\textb{$f_o T = \pm 10^{-4}$ could result from, e.g.,
oscillator error of $\pm 1$ ppm,
a carrier frequency of $1$~GHz, 
and bandwidth $1/T=10$~MHz.}}
$f_o T\sim\mc{U}[-10^{-4},-10^{-4})$,
where $\mc{U}[a,b)$ means ``uniformly distributed on the interval $[a,b)$.''
\color{black}
Unless otherwise noted, we used 
$M=64$ array elements,
$L=1024$ total symbols,
$Q=32$ training symbols, 
$N=5$ interferers,
\textb{and an oversampling factor of $P=2$.}
(Note that $Q\ll M$ \textb{but $Q\gg N$}.)

The quantities $\vec{h}$, $\vec{s}$, $\vec{B}$, and $\vec{\Phi}$ in \eqref{Ynum} were then constructed as follows.
The symbols in $\vec{s}$ were i.i.d.\ QPSK with variance $1$,
the noise $\vec{W}$ was i.i.d.\ circular Gaussian with variance $\nu$, and
the interference $\vec{\Phi}$ had entries with variance $\sigma_i^2/N$, 
giving a total interference power of $\sigma_i^2$.
\blue
Several types of interference $\vec{\Phi}$ were considered:
\begin{enumerate}
\item i.i.d.\ circular Gaussian,
\item unsynchronized QPSK, where 
$\vec{\Phi}=[\vec{\phi}_1,\dots,\vec{\phi}_N]$ with
$\vec{\phi}_n\herm = e^{j\theta_n}\vec{s}_n\herm \vec{G}_{\tilde{\tau}_n} \vec{J}_{f_n T}$
and $\theta_n\sim\mc{U}[0,2\pi)$, i.i.d.\ QPSK $\vec{s}_n$, $\tilde{\tau}_n\sim\mc{U}[-0.5,0.5)$, and $f_n T\sim\mc{U}[-10^{-4},-10^{-4})$,
\item sinusoidal, where
$\phi_n(t) = \sqrt{\sigma_i^2/N} e^{j(\omega_n t+\theta_n)}$
with $\theta_n\sim\mc{U}[0,2\pi)$ and $\omega_n\sim\mc{U}[-\pi/T,\pi/T)$, and
\item spike-like, where
$\phi_n(t) = \sqrt{\sigma_i^2 L/N} e^{j\theta_n}g(t-\tau_n T)$
with $\theta_n\sim\mc{U}[0,2\pi)$ and $\tau_n\sim\mc{U}[0,L)$.
\end{enumerate}
\color{black}
For the antenna array, we assumed a uniform planar array (UPA) with half-wavelength element spacing operating in the narrowband regime.
Then, to generate the signal's array response $\vec{h}$, we assumed that the signal arrived from a random (horizontal,vertical) angle pair drawn uniformly on $[0,2\pi)^2$. 
For the $n$th interferer's array response $\vec{b}_n$, we used the arrival angle corresponding to the $n$th largest sidelobe in $\vec{h}$.

\newcommand{\ket}{\textsf{kel-tr}\xspace}
\newcommand{\gst}{\textsf{kmr-tr}\xspace}
\newcommand{\mct}{\textsf{mcw-tr}\xspace}
\newcommand{\kez}{\textsf{kel-1}\xspace}
\newcommand{\gsz}{\textsf{kmr-1}\xspace}
\newcommand{\mcz}{\textsf{mcw-1}\xspace}
\newcommand{\kee}{\textsf{kel-em}\xspace}
\newcommand{\gse}{\textsf{kmr-em}\xspace}
\newcommand{\mce}{\textsf{mcw-em}\xspace}
\newcommand{\keeh}{\textsf{forsythe}\xspace}
\newcommand{\gseh}{\textsf{forsythe-lowrank}\xspace}
\newcommand{\mceh}{\textsf{hard-mcw-em}\xspace}

The following detectors were tested.
First, we considered several existing methods that used only the training data $\vec{Y}\train$:
\begin{enumerate}
\item
the Kang/Monga/Rangaswamy (KMR) approach \eqref{glrt_kmr2}, \textb{but with interference rank $N$ estimated\footnote{%
\textb{We emphasize that, whereas the original KMR \cite{Kang:TAES:14} and McWhorter \cite{McWhorter:ASAP:04} detectors assume \emph{known} interference rank $N$, we simulate enhanced versions of these detections that \emph{estimate} $N$.  We do this to meaningfully compare to the proposed detectors, which also estimate $N$.
Over our suite of experiments, we found that \gst worked well with the GIC rule from \eqref{GIC} under $G=1.1$, and \mct worked well with the GIC rule under $G=1.25$.
}
} as described in \secref{rankGS}, i.e., ``\gst.''}
\item
McWhorter's approach \eqref{glrt_mc2}, \textb{but with interference rank $N$ estimated as described in \secref{rankMc}}, i.e., ``\mct.''
\item
Kelly's full-rank approach \eqref{glrt_kelly2}, i.e., ``\ket.''
\end{enumerate}
We also tested the proposed EM-based methods, which use the full data $\vec{Y}$.
In particular, we tested 
\begin{enumerate}
\item
\algref{GS} with $N$ estimated as in \secref{rankGS}, i.e., ``\gse.''
\item
\algref{McW} with $N$ estimated as in \secref{rankMc}, i.e., ``\mce''
\item
\algref{GS} with full rank $N=M$, i.e., ``\kee.'' 
\end{enumerate}
For the EM algorithm, we used a maximum of $50$ iterations but terminated early, at iteration $i>1$, if $\|\hvec{s}\of{i}-\hvec{s}\of{i-1}\|/\|\hvec{s}\of{i}\|<0.01$.

We also tested Forsythe's iterative method \cite[p.~110]{Forsythe:LLJ:97} by running 
\algref{GS} with full rank $N=M$ and hard symbol estimates in lines~\ref{line:shat}-\ref{line:E}, 
as discussed in \secref{Forsythe}.
In addition, we tested a low-rank version of Forsythe's method by running 
\algref{GS} with hard estimates and $N$ estimated as in \secref{rankGS}. 
Finally, we tested \algref{McW} with hard estimates and $N$ estimated as in \secref{rankMc} (denoted by ``\mceh'').

For all methods, detection performance was quantified using 
\iftoggle{pd}{
the rate of correct detection when the detector threshold $\eta$ is set to achieve a fixed false-alarm rate.
}{
\begin{align}
\Pr\{\text{error}\} &= \Pr\{\text{missed detection}\} + \Pr\{\text{false alarm}\}
\end{align}
after setting each detector's threshold $\eta$ to balance the miss and false-alarm rates. 
Note that this is the error-rate minimizing choice of $\eta$ \cite{Scharf:Book:91}.
}
All simulation results represent the average of $10\,000$ independent draws of $\{\vec{h},\vec{s},\vec{B},\vec{\Phi},\vec{W},\textb{\tilde{\tau}_o,f_o}\}$.

\blue
\subsection{Performance versus timing synchronization error}

\iftoggle{pd}{\Figref{demo6_iG_pd_pfa001}}{\Figref{}}
shows \metric versus\footnote{\textb{In this experiment, $\tilde{\tau}_o$ was fixed, while in all other experiments $\tilde{\tau}_o$ was randomly drawn from the distribution $\mc{U}[-\tfrac{1}{2P},\tfrac{1}{2P}]$}.} 
baud-normalized timing synchronization error $\tilde{\tau}_o$ for various detectors under $\nu=\sigma_i^2 = Q$ 
and i.i.d.\ Gaussian interference.
There we see that all methods degrade as $\tilde{\tau}_o$ increases, but that the proposed low-rank, EM-based methods \gse and \mce outperform the others.
We also see that timing offsets $|\tilde{\tau}_o|<0.25$ have a negligible effect on \gse and \mce, a small effect on the low-rank detectors \gst and \mct, and a larger effect on the full-rank detectors \kee and \ket.

\iftoggle{pd}{\putFrag{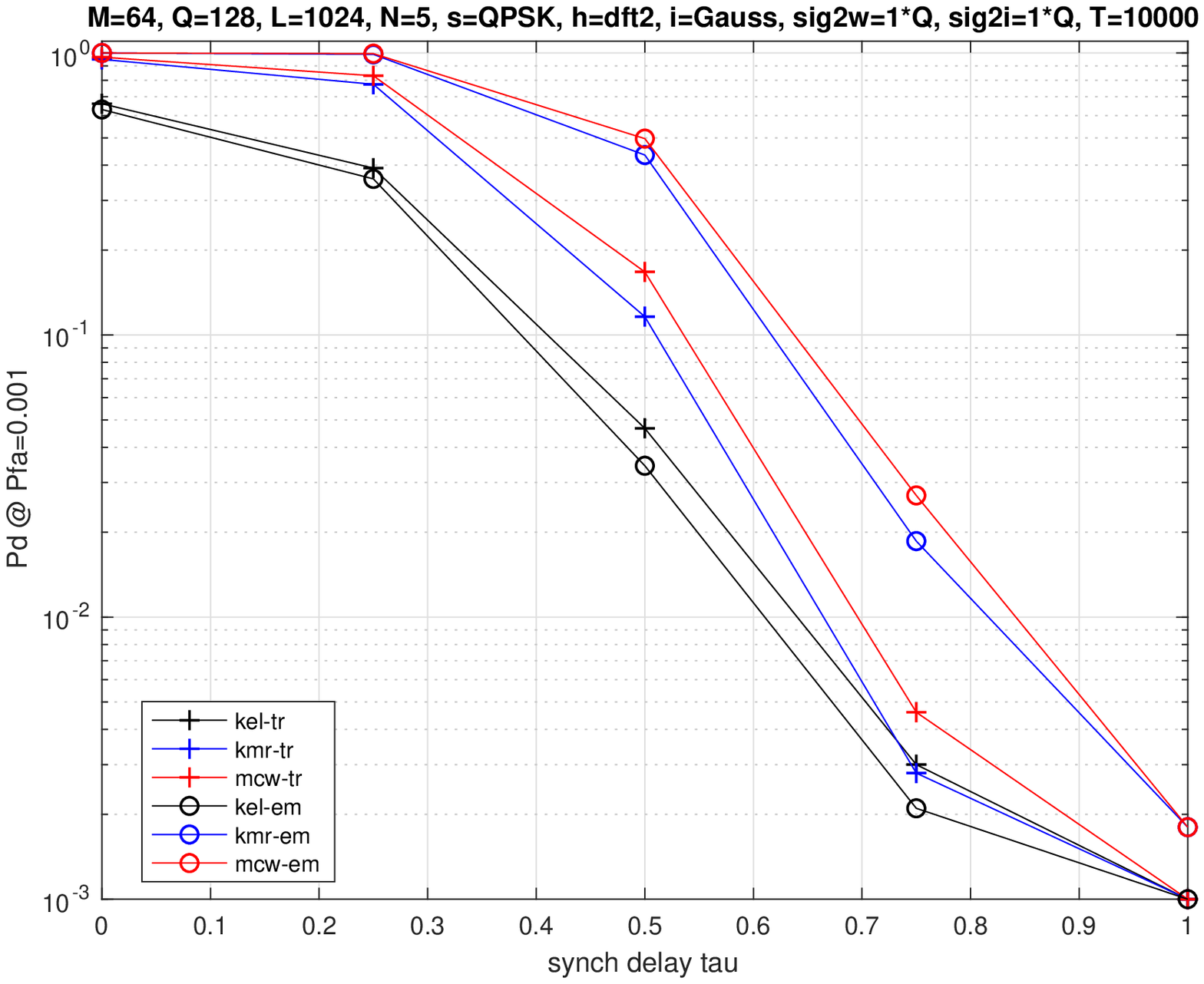}}{\putFrag{}}
        {\textb{\Metric versus $\tilde{\tau}_o$ for various detectors, under
         $\nu=\sigma_i^2=Q$,
         $M=64$, $Q=128$, $L=1024$, $N=5$, i.i.d.\ QPSK symbols, and
         \textb{i.i.d.\ Gaussian interference}. 
         The proposed low-rank, EM-based methods \gse and \mce 
         are robust to timing offsets $|\tilde{\tau}_o|<0.5$ 
         and perform better than the other methods.}}
        {\figsize}
        {\psfrag{Pr(err)}[B][B][0.8]{\sf $\Pr\{\text{error}\}$} 
         \psfrag{Pd @ Pfa=0.001}[B][B][0.8]{\sf \metric} 
         \psfrag{synch delay tau}[t][t][0.8]{$\tilde{\tau}_o$}
         \psfrag{M=64, Q=128, L=1024, N=5, s=QPSK, h=dft2, i=Gauss, sig2w=1*Q, sig2i=1*Q, T=10000}{}
         }

\color{black}
\subsection{Performance versus training length \texorpdfstring{$Q$}{Q}}

\iftoggle{pd}{\Figref{demo1_iG_pd_pfa001}}{\Figref{demo1_10k_perr}}
shows \metric versus training length $Q$ for various detectors under $\nu=\sigma_i^2=Q$ \textb{and i.i.d.\ Gaussian interference}.
Here, $\nu$ and $\sigma_i^2$ grow with $Q$ to prevent the error-rate from vanishing with $Q$ due to spreading gain.
The \ket trace is clipped on the left because Kelly's approach is not defined when $Q<M$. 
\iftoggle{pd}{\Figref{demo1_iG_pd_pfa001}}{\Figref{demo1_10k_perr}}
shows that the proposed EM-based, low-rank detectors \gse and \mce outperformed the others for $Q\in[16,256]$.
For $Q=512$, \gse and \mce performed on par with \gst and \mct.
When $Q=1024=L$, there are no data symbols, and so \gse and \mce are equivalent to \gst and \mct.

\iftoggle{pd}{\putFrag{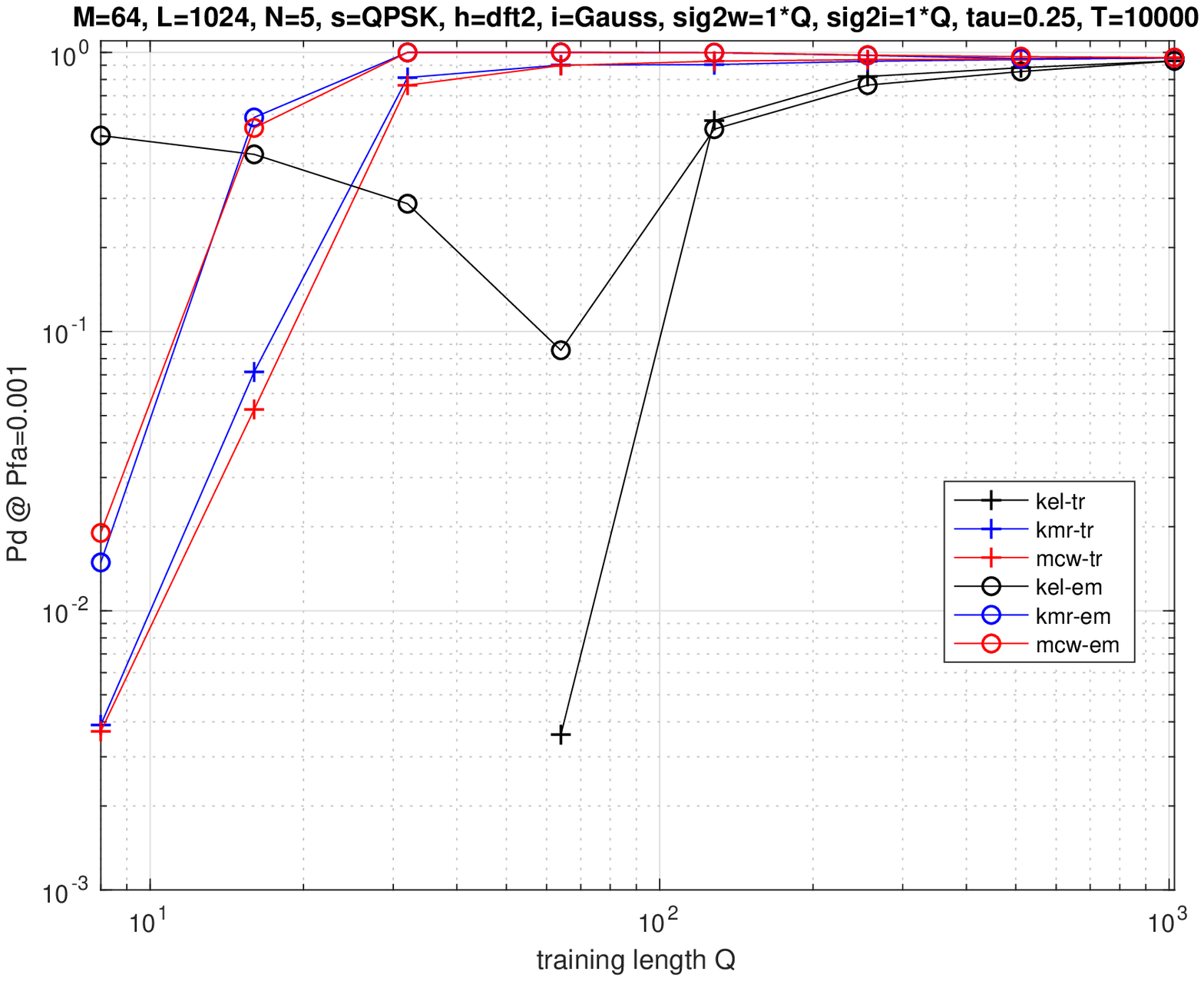}}{\putFrag{demo1_10k_perr}}
        {\Metric versus training length $Q$ for various detectors, under
         $\nu=\sigma_i^2=Q$,
         $M=64$, $L=1024$, $N=5$, i.i.d.\ QPSK symbols, and
         \textb{i.i.d.\ Gaussian interference}. 
         The proposed EM-based, low-rank detectors \gse and \mce 
         \iftoggle{pd}{outperform the others for $Q\in[16,256]$}
         {perform far better than the others for $Q\in[32,256]$}.}
        {\figsize}
        {\psfrag{Pr(err)}[B][B][0.8]{\sf $\Pr\{\text{error}\}$} 
         \psfrag{Pd @ Pfa=0.001}[B][B][0.8]{\sf \metric} 
         \psfrag{training length Q}[t][t][0.8]{\sf training length $Q$} 
         \psfrag{M=64, L=1024, N=5, s=QPSK, h=dft2, i=Gauss, sig2w=1*Q, sig2i=1*Q, tau=0.25, T=10000}{}}

\subsection{Performance versus SINR}

\iftoggle{pd}{\Figref{demo2_iG_pd_pfa001}}{\Figref{demo2_10k_perr}}
shows \metric versus $\nu=\sigma_i^2$ for various detectors \textb{under i.i.d.\ Gaussian interference}. 
\textb{For this and subsequent experiments, we focus on the challenging case where the number of training symbols, $Q=32$, is only half of the number of antennas, $M=64$, in which case the \ket method is undefined.
Consequently, results for \ket are not shown.}
In \iftoggle{pd}{\Figref{demo2_iG_pd_pfa001}}{\Figref{demo2_10k_perr}}
we see that the proposed EM-based, full-data detectors \gse and \mce significantly outperformed their training-based counterparts \gst and \mct.

\iftoggle{pd}{\putFrag{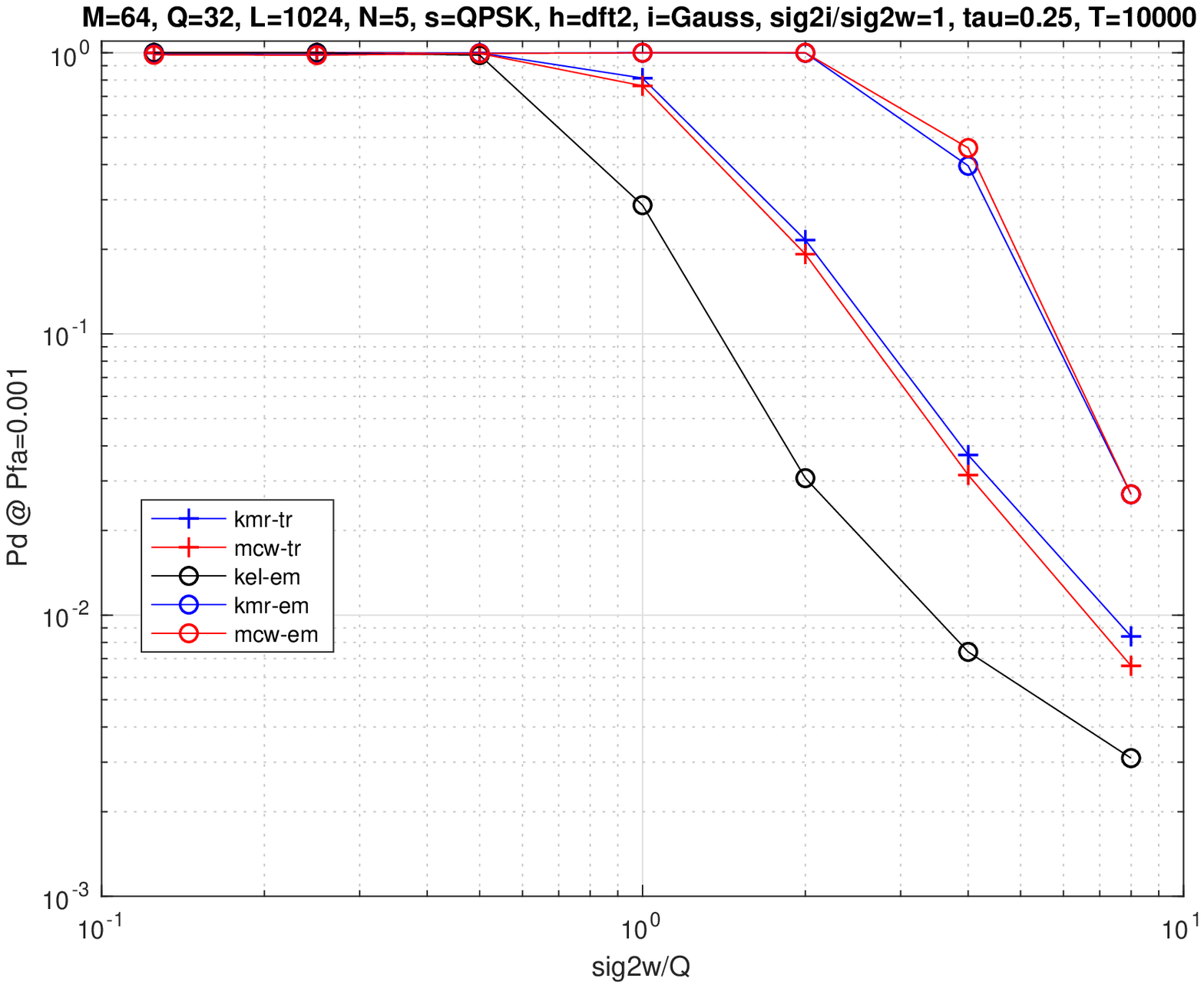}}{\putFrag{demo2_10k_perr}}
        {\Metric versus $\nu=\sigma_i^2$ for various detectors, under
         $M=64$, $Q=32$, $L=1024$, $N=5$, i.i.d.\ QPSK symbols, and
         \textb{i.i.d.\ Gaussian interference}. 
         The proposed low-rank, EM-based methods \gse and \mce perform far better than the others.}
        {\figsize}
        {\psfrag{Pr(err)}[B][B][0.8]{\sf $\Pr\{\text{error}\}$} 
         \psfrag{Pd @ Pfa=0.001}[B][B][0.8]{\sf \metric} 
         \psfrag{sig2w/Q}[t][t][0.8]{$\nu/Q = \sigma_i^2/Q$}
         \psfrag{M=64, Q=32, L=1024, N=5, s=QPSK, h=dft2, i=Gauss, sig2i/sig2w=1, tau=0.25, T=10000}{}
         }

\iftoggle{pd}{\Figref{demo2_iG_pd_pfa001_hard}}{\Figref{demo2_10k_hard_perr}}
shows the performance of Forsythe's full-rank iterative method, its low-rank counterpart (i.e., \algref{GS} with hard symbol estimates), and \algref{McW} with hard symbol estimates, under the same data used to create 
\iftoggle{pd}{\figref{demo2_iG_pd_pfa001}}{\figref{demo2_10k_perr}}.
Comparing the two figures, we see that the ``soft'' 
methods, \kee, \gse, and \mce, outperformed their hard counterparts, \keeh, \gseh, and \mceh.
We attribute this behavior to error propagation in the hard detector. 
Also, we see that the low-rank methods outperformed the full-rank methods, which is expected since the interference is truly of low rank.

\iftoggle{pd}{\putFrag{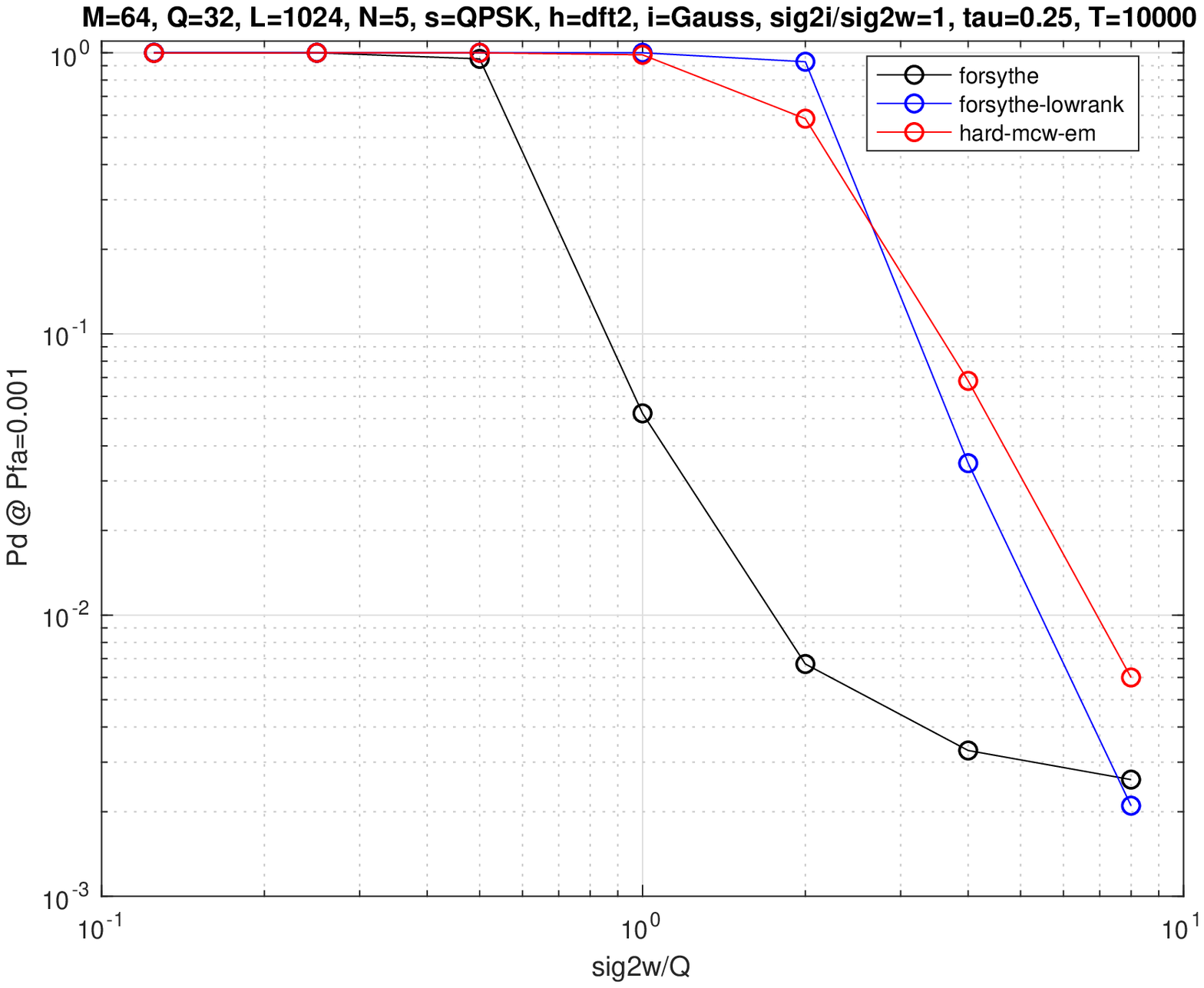}}{\putFrag{demo2_10k_hard_perr}}
        {\Metric versus $\nu=\sigma_i^2$ for various ``hard'' symbol detectors, under
         $M=64$, $Q=32$, $L=1024$, $N=5$, i.i.d.\ QPSK symbols, and
         \textb{i.i.d.\ Gaussian interference}. 
         Comparing to \iftoggle{pd}{\figref{demo2_iG_pd_pfa001}}{\figref{demo2_10k_perr}}, these hard detectors do not perform as well as the proposed ``soft'' detectors \gse and \mce.}
        {\figsize}
        {\psfrag{Pr(err)}[B][B][0.8]{\sf $\Pr\{\text{error}\}$} 
         \psfrag{Pd @ Pfa=0.001}[B][B][0.8]{\sf \metric} 
         \psfrag{sig2w/Q}[t][t][0.8]{$\nu/Q = \sigma_i^2/Q$}
         \psfrag{M=64, Q=32, L=1024, N=5, s=QPSK, h=dft2, i=Gauss, sig2i/sig2w=1, tau=0.25, T=10000}{}}

\subsection{Performance versus SIR at fixed SNR}

\renewcommand{\pfa}{$10^{-2}$}

\iftoggle{pd}{\Figref{demo3_iG_pd_pfa01}}{\Figref{demo3_10k_perr}}
shows \metric versus interference power $\sigma_i^2$ at the fixed noise power $\nu=Q$.
\textb{In this experiment, the interference was i.i.d.\ Gaussian.}
The proposed EM-based, low-rank detectors \gse and \mce gave no errors over $10\,000$ trials%
\iftoggle{pd}{.}{, and thus the \gse and \mce traces are not visible in \figref{demo3_10k_perr}.}
In fact, \gse and \mce remained error-free for arbitrarily large $\sigma_i^2$, suggesting that they correctly learned the interference subspace and avoided it completely.
The non-monotonic behavior of the training based schemes, \gst and \mct,
results from imperfect rank estimation:
when $\sigma_i^2\gg \nu$ the rank was estimated as $\hat{N}=N$, 
and when $\sigma_i^2\ll \nu$ the rank was estimated as $\hat{N}=0$, 
but when $\sigma_i^2 \approx \nu$ it was difficult to estimate the rank, leading to detection errors.


\iftoggle{pd}{\putFrag{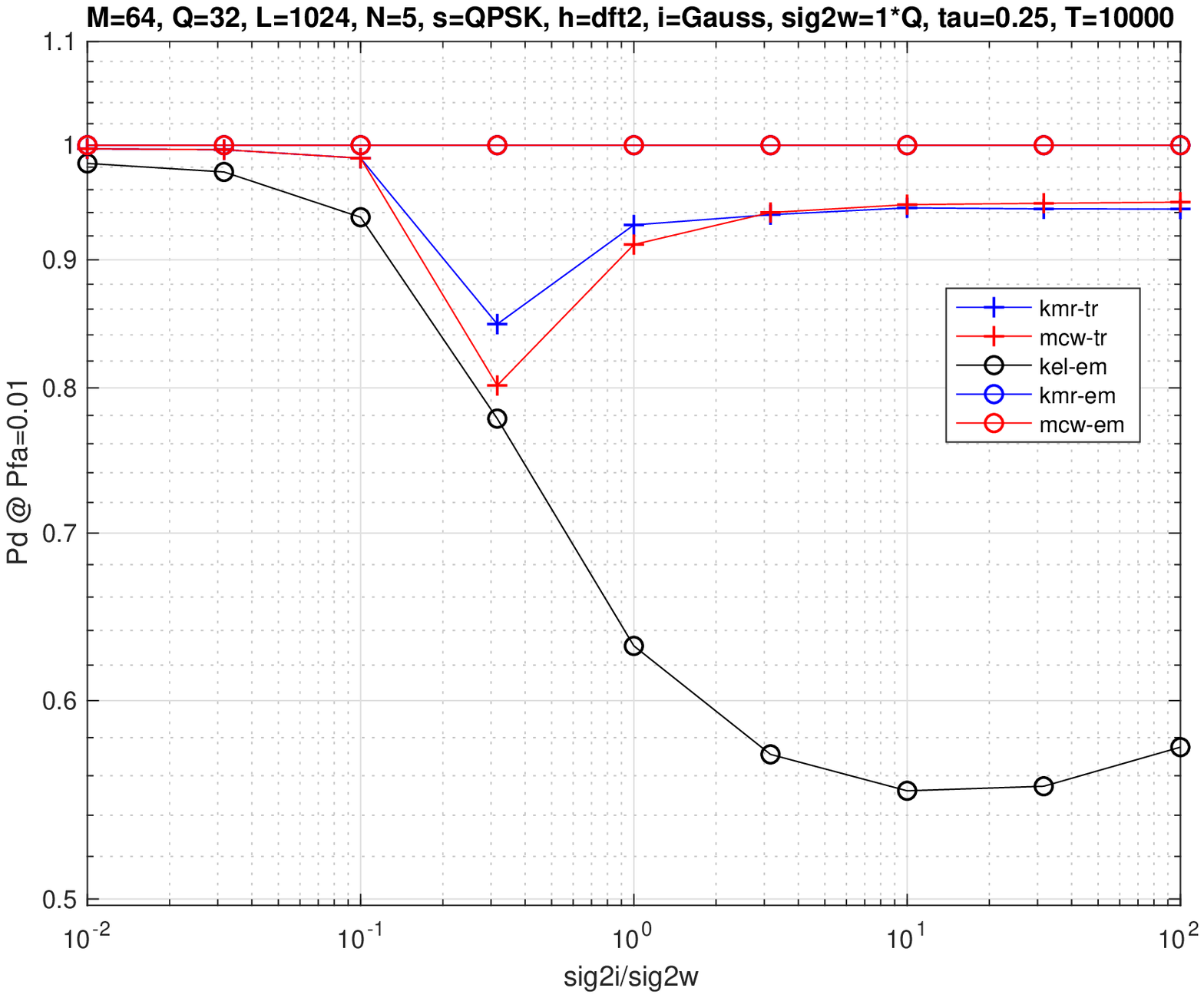}}{\putFrag{demo3_10k_perr}}
        {\Metric versus $\sigma_i^2$ for various detectors, under 
         $\nu=Q$,
         $M=64$, $Q=32$, $L=1024$, $N=5$, i.i.d.\ QPSK symbols, and
         \textb{i.i.d.\ Gaussian interference}. 
         The proposed EM-based, low-rank methods \gse and \mce gave zero errors over $10\,000$ realizations\iftoggle{pd}{.}{ and thus are not visible in the figure.}}
        {\figsize}
        {\psfrag{Pr(err)}[B][B][0.8]{\sf $\Pr\{\text{error}\}$} 
         \psfrag{Pd @ Pfa=0.01}[B][B][0.8]{\sf \metric} 
         \psfrag{sig2i/sig2w}[t][t][0.8]{$\sigma_i^2/Q$}
         \psfrag{M=64, Q=32, L=1024, N=5, s=QPSK, h=dft2, i=Gauss, sig2w=1*Q, tau=0.25, T=10000}{}}

\blue
\iftoggle{pd}{\Figref{demo3_iQ_pd_pfa01}}{\Figref{}}
repeats the experiment, but with unsynchronized QPSK interference, constructed as described in \secref{setup}.
Qualitatively, the results are similar to the case of i.i.d.\ Gaussian interference.

\iftoggle{pd}{\putFrag{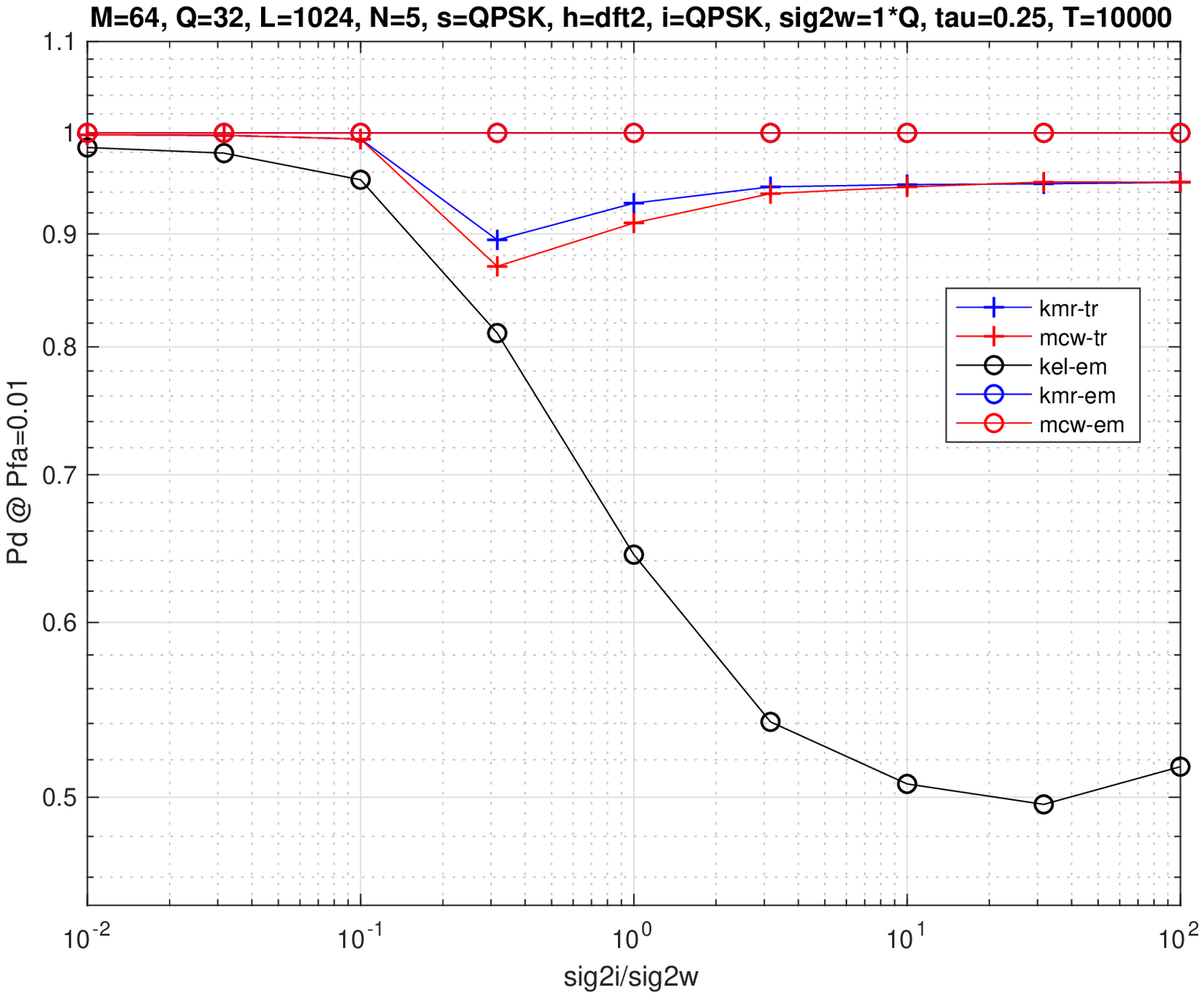}}{\putFrag{}}
        {\textb{\Metric versus $\sigma_i^2$ for various detectors, under 
         $\nu=Q$,
         $M=64$, $Q=32$, $L=1024$, $N=5$, i.i.d.\ QPSK symbols, and
         unsynchronized QPSK interference.}}
        {\figsize}
        {\psfrag{Pr(err)}[B][B][0.8]{\sf $\Pr\{\text{error}\}$} 
         \psfrag{Pd @ Pfa=0.01}[B][B][0.8]{\sf \metric} 
         \psfrag{sig2i/sig2w}[t][t][0.8]{$\sigma_i^2/Q$}
         \psfrag{M=64, Q=32, L=1024, N=5, s=QPSK, h=dft2, i=QPSK, sig2w=1*Q, tau=0.25, T=10000}{}}

\iftoggle{pd}{\Figref{demo3_iS_pd_pfa01}}{\Figref{}}
repeats the experiment, but with sinusoidal interference. 
The results are similar, except that \kee performs worse when the interference is very strong.

\iftoggle{pd}{\putFrag{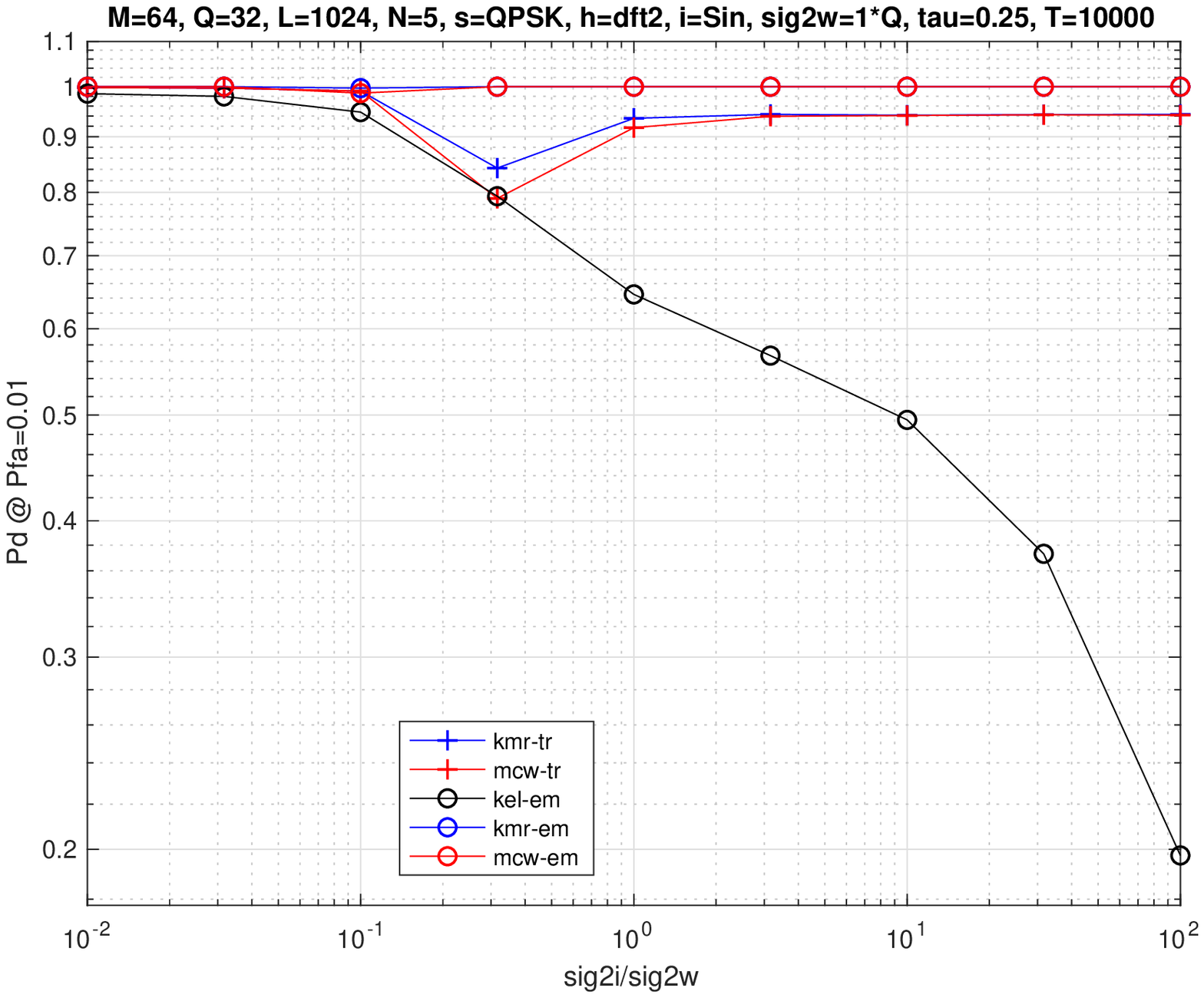}}{\putFrag{}}
        {\textb{\Metric versus $\sigma_i^2$ for various detectors, under 
         $\nu=Q$,
         $M=64$, $Q=32$, $L=1024$, $N=5$, i.i.d.\ QPSK symbols, and
         sinusoidal interference.}}
        {\figsize}
        {\psfrag{Pr(err)}[B][B][0.8]{\sf $\Pr\{\text{error}\}$} 
         \psfrag{Pd @ Pfa=0.01}[B][B][0.8]{\sf \metric} 
         \psfrag{sig2i/sig2w}[t][t][0.8]{$\sigma_i^2/Q$}
         \psfrag{M=64, Q=32, L=1024, N=5, s=QPSK, h=dft2, i=Sin, sig2w=1*Q, tau=0.25, T=10000}{}}

\iftoggle{pd}{\Figref{demo3_iD_pd_pfa01}}{\Figref{}}
repeats the experiment, but with spike-like interference. 
All detectors find the spike-like interference much easier to handle than
i.i.d.\ Gaussian, QPSK, and sinusoidal interference. 

\iftoggle{pd}{\putFrag{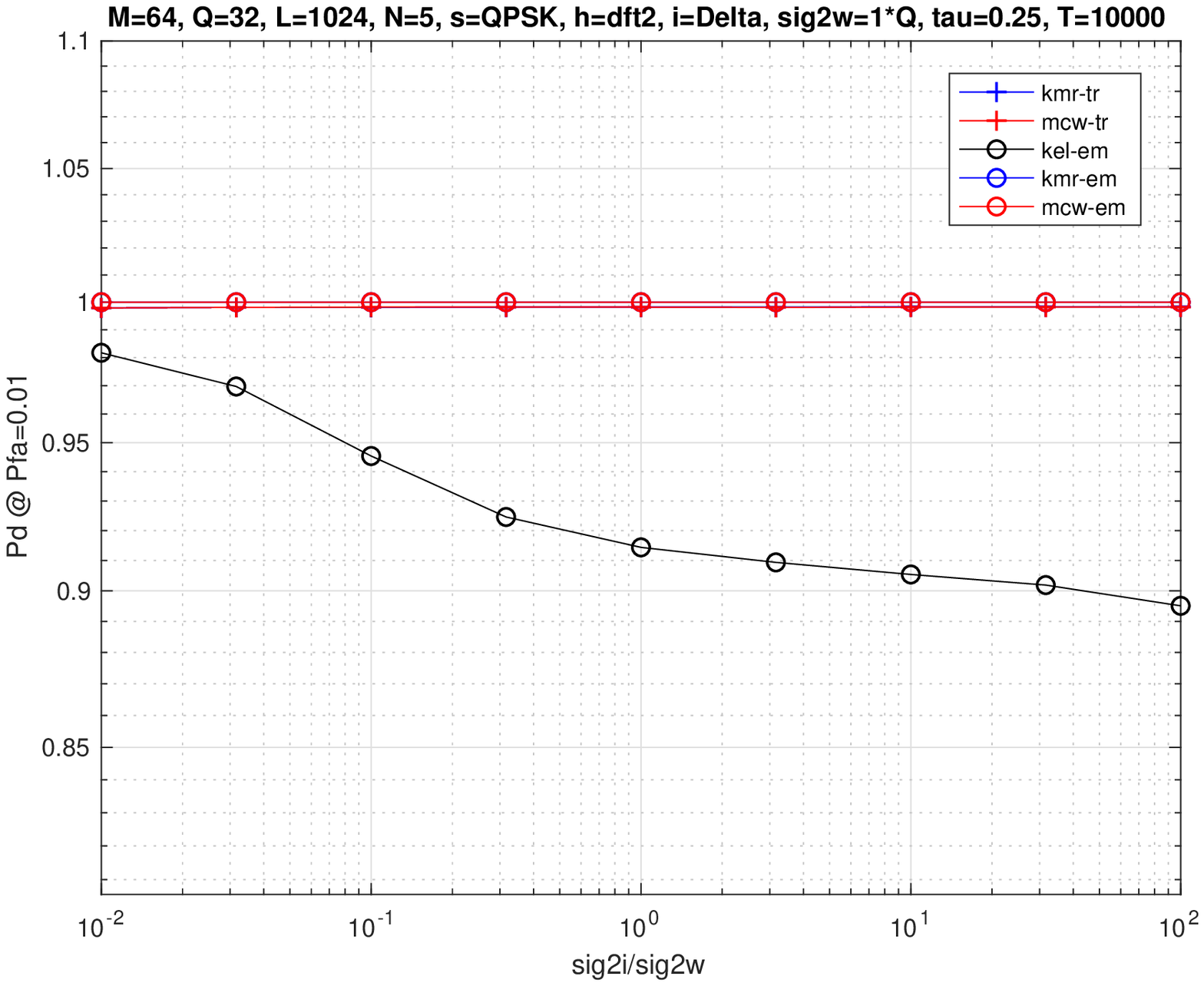}}{\putFrag{}}
        {\textb{\Metric versus $\sigma_i^2$ for various detectors, under 
         $\nu=Q$,
         $M=64$, $Q=32$, $L=1024$, $N=5$, i.i.d.\ QPSK symbols, and
         spike-like interference.}}
        {\figsize}
        {\psfrag{Pr(err)}[B][B][0.8]{\sf $\Pr\{\text{error}\}$} 
         \psfrag{Pd @ Pfa=0.01}[B][B][0.8]{\sf \metric} 
         \psfrag{sig2i/sig2w}[t][t][0.8]{$\sigma_i^2/Q$}
         \psfrag{M=64, Q=32, L=1024, N=5, s=QPSK, h=dft2, i=Delta, sig2w=1*Q, tau=0.25, T=10000}{}}

\color{black}

\subsection{Performance versus interference rank \texorpdfstring{$N$}{N}}

\iftoggle{pd}{\Figref{demo4_iG_pd_pfa01}}{\Figref{demo4_10k_perr}}
shows \metric versus the number of interferers, $N$, for various detectors under $\nu=Q$ and $\sigma_i^2 = QN$.
Note that the per-interferer power was fixed at $Q$.
Note also that the proposed EM-based, low-rank detectors gave no errors over $10\,000$ trials\iftoggle{pd}{.}{, and so the \gse and \mce traces are not visible in \figref{demo4_10k_perr}.}
For the other schemes, the error-rate increased\iftoggle{pd}{}{ monotonically} with $N$, as expected.

\iftoggle{pd}{\putFrag{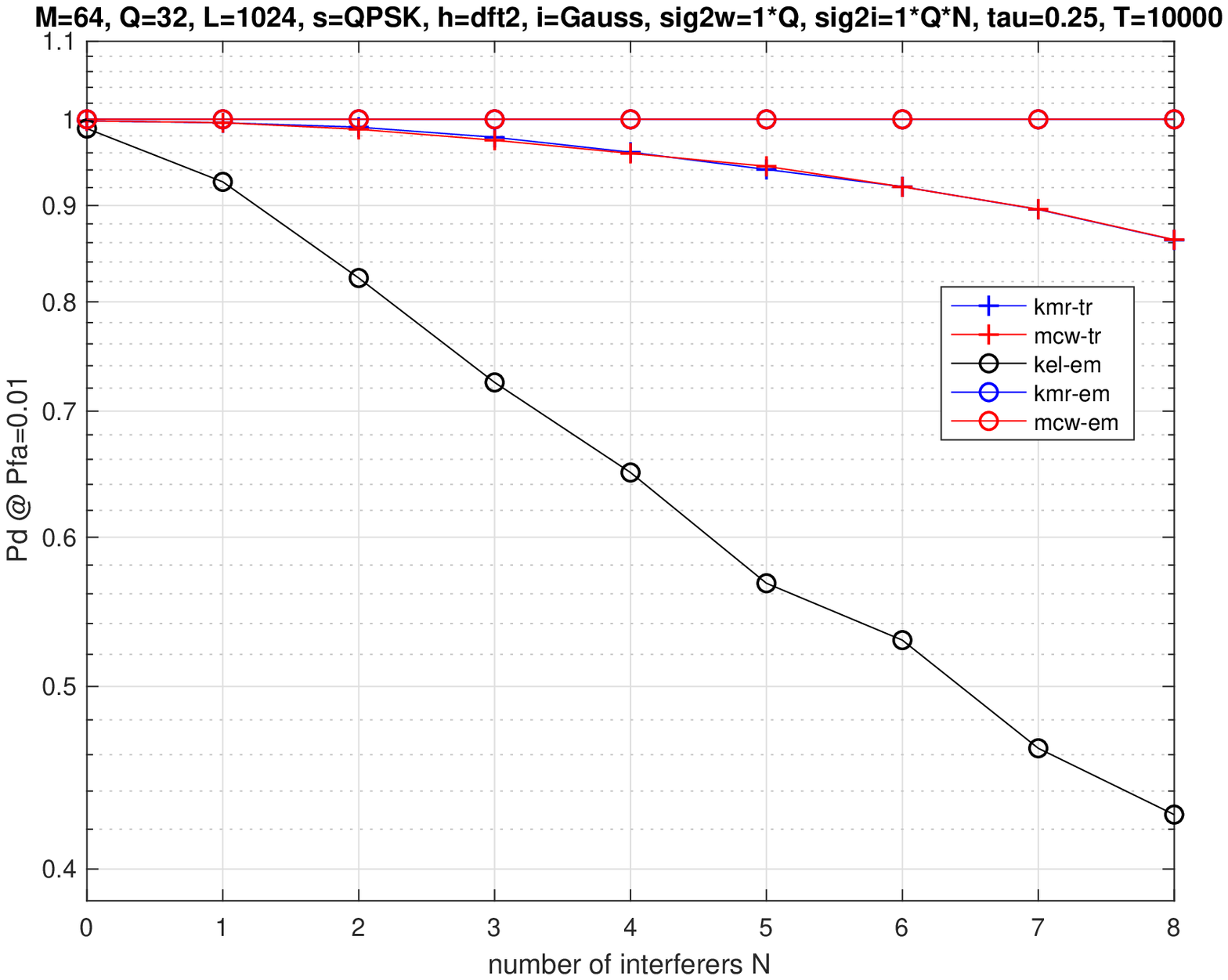}}{\putFrag{demo4_10k_perr}}
        {\Metric versus number of interferers $N$ for various detectors, under
         $\nu=Q$, $\sigma_i^2=QN$,
         $M=64$, $Q=32$, $L=1024$, $N=5$, i.i.d.\ QPSK symbols, and
         \textb{i.i.d.\ Gaussian interference}. 
         The proposed EM-based, low-rank detectors \gse and \mce gave zero errors over $10\,000$ realizations\iftoggle{pd}{.}{ and thus are not visible in the figure.}}
        {\figsize}
        {\psfrag{Pr(err)}[B][B][0.8]{\sf $\Pr\{\text{error}\}$} 
         \psfrag{Pd @ Pfa=0.01}[B][B][0.8]{\sf \metric} 
         \psfrag{number of interferers N}[t][t][0.8]{\sf number of interferers $N$} 
         \psfrag{M=64, Q=32, L=1024, s=QPSK, h=dft2, i=Gauss, sig2w=1*Q, sig2i=1*Q*N, tau=0.25, T=10000}{}}

\Figref{demo4_iG_Nhat} shows the average estimated interference rank $\hat{N}$ versus the true rank $N$ under $\mc{H}_1$, using the same data used to construct 
\iftoggle{pd}{\figref{demo4_iG_pd_pfa01}}{\figref{demo4_10k_perr}}.
There we see that all methods were successful, on average, at correctly estimating the interference rank.

\putFrag{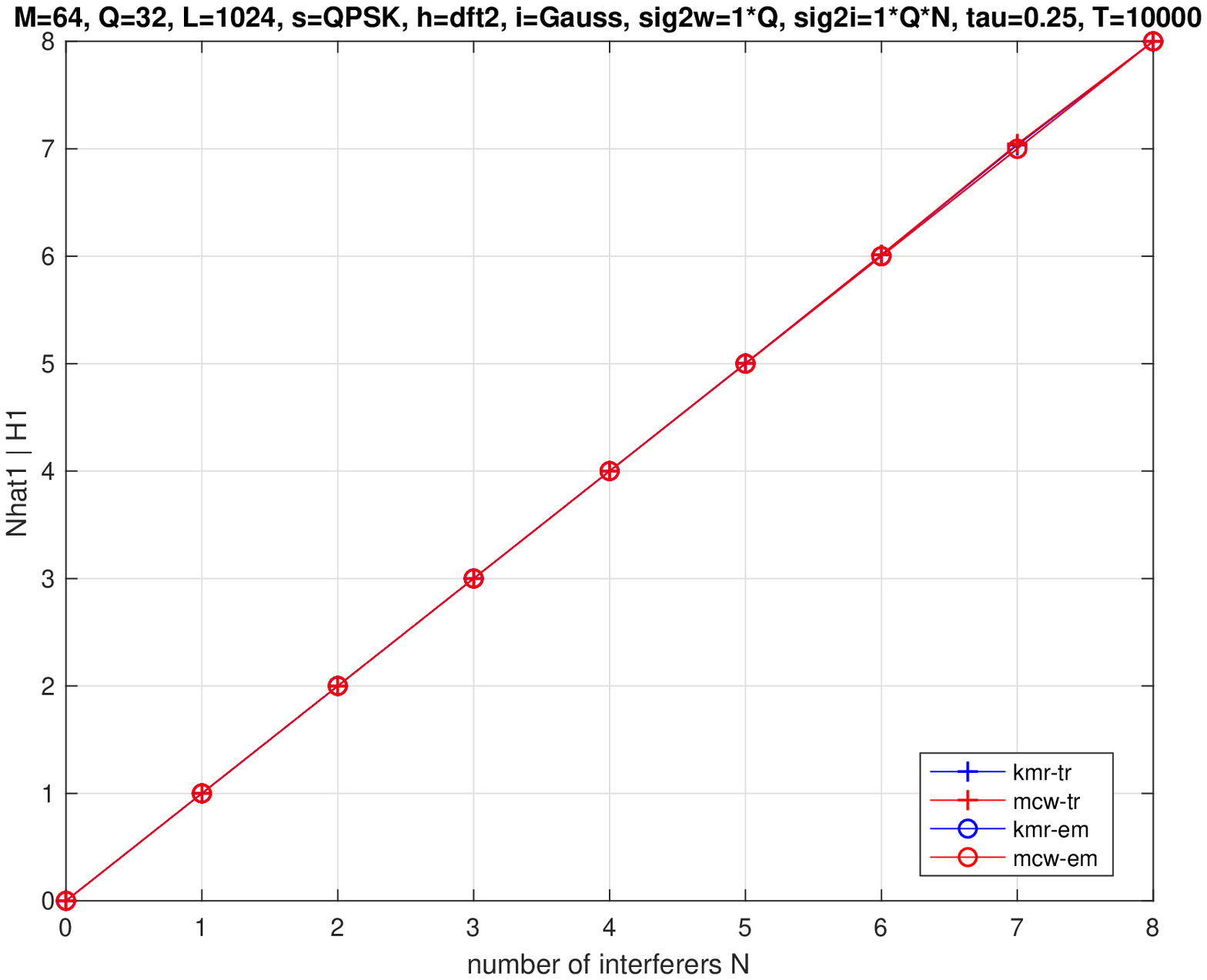}
        {Average rank estimate $\hat{N}$ versus true rank $N$ for various low-rank detectors, under
         $\mc{H}_1$, $\nu=Q$, $\sigma_i^2=QN$,
         $M=64$, $Q=32$, $L=1024$, i.i.d.\ QPSK symbols, and
         \textb{i.i.d.\ Gaussian interference}. 
         For all detectors, $\hat{N}\approx N$.}
        {\figsize}
        {\psfrag{Nhat1 | H1}[b][b][0.8]{\sf estimated interference rank $\hat{N}$ under $\mc{H}_1$}  
         \psfrag{number of interferers N}[t][t][0.8]{\sf true interference rank $N$} 
         \iftoggle{pd}{\psfrag{M=64, Q=32, L=1024, s=QPSK, h=dft2, sig2w=1*Q, sig2i=1*Q*N, T=10000}{}}{}
         \psfrag{M=64, Q=32, L=1024, s=QPSK, h=dft2, i=Gauss, sig2w=1*Q, sig2i=1*Q*N, tau=0.25, T=10000}{}
         }

\blue
We now repeat the experiment that generated 
\iftoggle{pd}{\Figref{demo4_iG_pd_pfa01}}{\Figref{demo4_10k_perr}},
but now using unsynchronized QPSK interference.
\iftoggle{pd}{\Figref{demo4_iQ_pd_pfa01}}{\Figref{}}
shows that the results are very similar.
We then repeat the same experiment again, but 
with sinusoidal interference.
\iftoggle{pd}{\Figref{demo4_iS_pd_pfa01}}{\Figref{}}
shows that the results are again quite similar.
Finally, we repeat the experiment 
with spike-like interference.
\iftoggle{pd}{\Figref{demo4_iD_pd_pfa01}}{\Figref{}}
shows that spike-like interference is much easier to handle than
i.i.d.\ Gaussian, QPSK, and sinusoidal interference.

\iftoggle{pd}{\putFrag{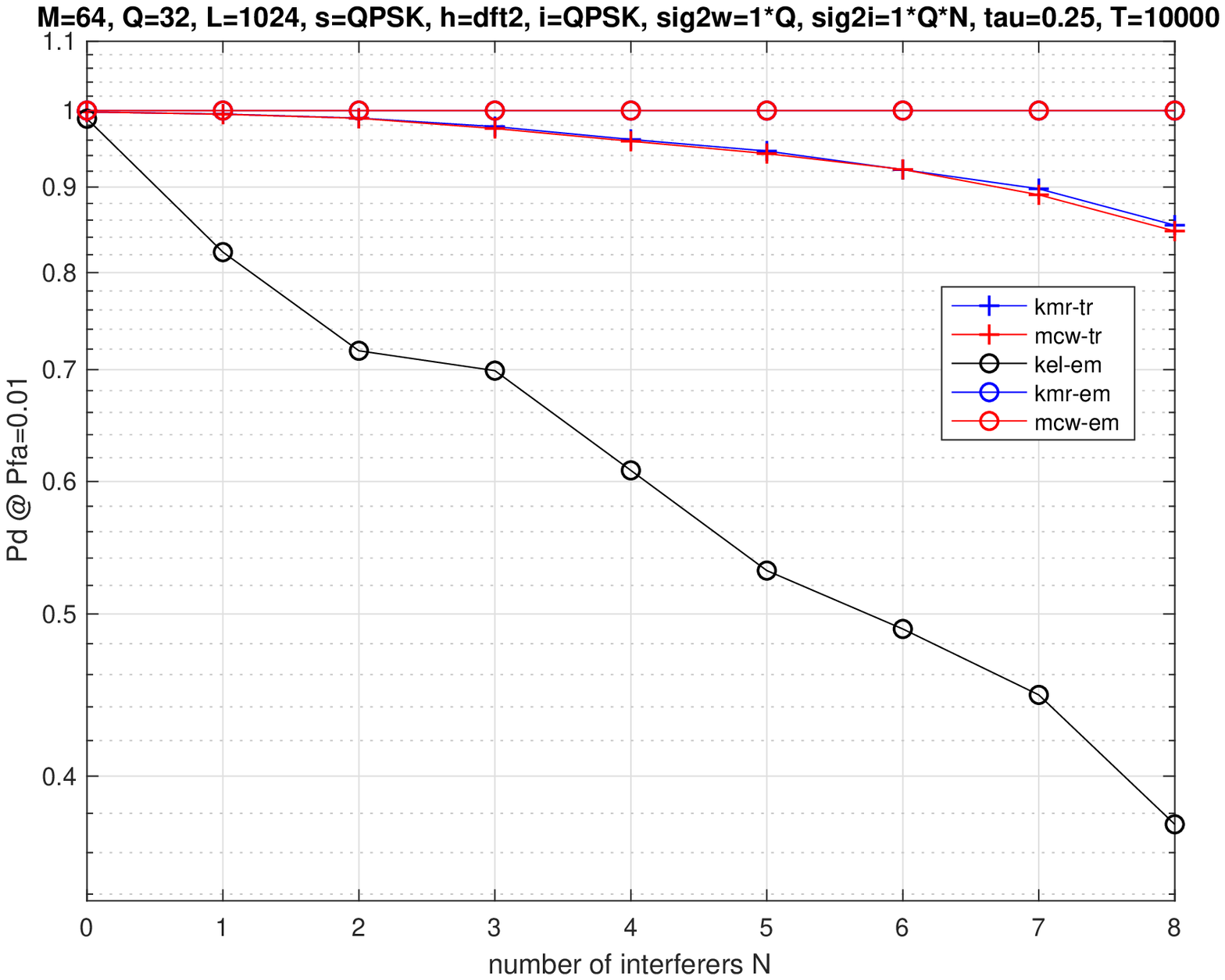}}{\putFrag{}}
        {\textb{\Metric versus number of interferers $N$ for various detectors, under
         $\nu=Q$, $\sigma_i^2=QN$,
         $M=64$, $Q=32$, $L=1024$, $N=5$, i.i.d.\ QPSK symbols, and
         unsynchronized QPSK interference.}}
        {\figsize}
        {\psfrag{Pr(err)}[B][B][0.8]{\sf $\Pr\{\text{error}\}$} 
         \psfrag{Pd @ Pfa=0.01}[B][B][0.8]{\sf \metric} 
         \psfrag{number of interferers N}[t][t][0.8]{\sf number of interferers $N$} 
         \psfrag{M=64, Q=32, L=1024, s=QPSK, h=dft2, i=QPSK, sig2w=1*Q, sig2i=1*Q*N, tau=0.25, T=10000}{}}

\iftoggle{pd}{\putFrag{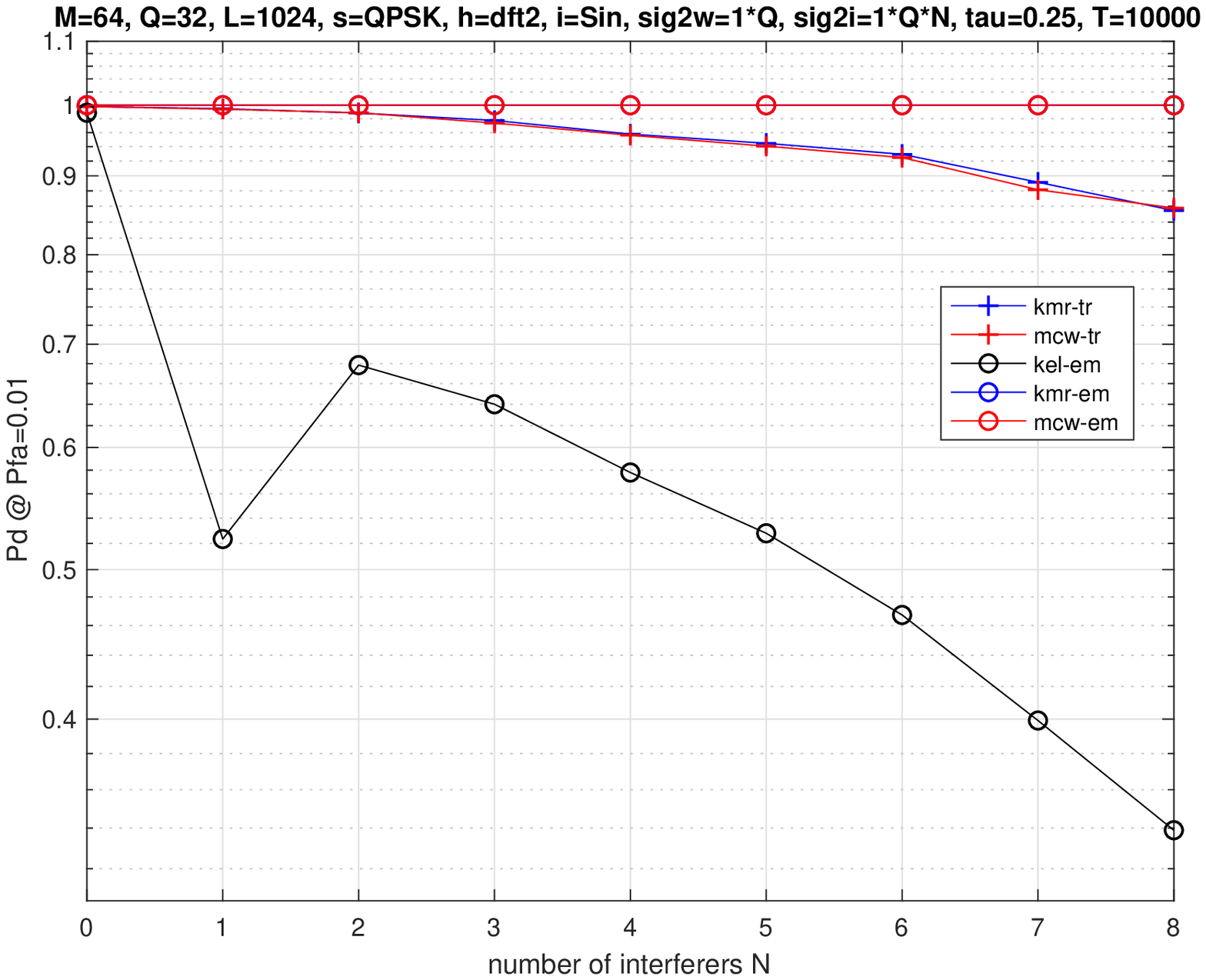}}{\putFrag{}}
        {\textb{\Metric versus number of interferers $N$ for various detectors, under
         $\nu=Q$, $\sigma_i^2=QN$,
         $M=64$, $Q=32$, $L=1024$, $N=5$, i.i.d.\ QPSK symbols, and
         sinusoidal interference.}}
        {\figsize}
        {\psfrag{Pr(err)}[B][B][0.8]{\sf $\Pr\{\text{error}\}$} 
         \psfrag{Pd @ Pfa=0.01}[B][B][0.8]{\sf \metric} 
         \psfrag{number of interferers N}[t][t][0.8]{\sf number of interferers $N$} 
         \psfrag{M=64, Q=32, L=1024, s=QPSK, h=dft2, i=Sin, sig2w=1*Q, sig2i=1*Q*N, tau=0.25, T=10000}{}}

\iftoggle{pd}{\putFrag{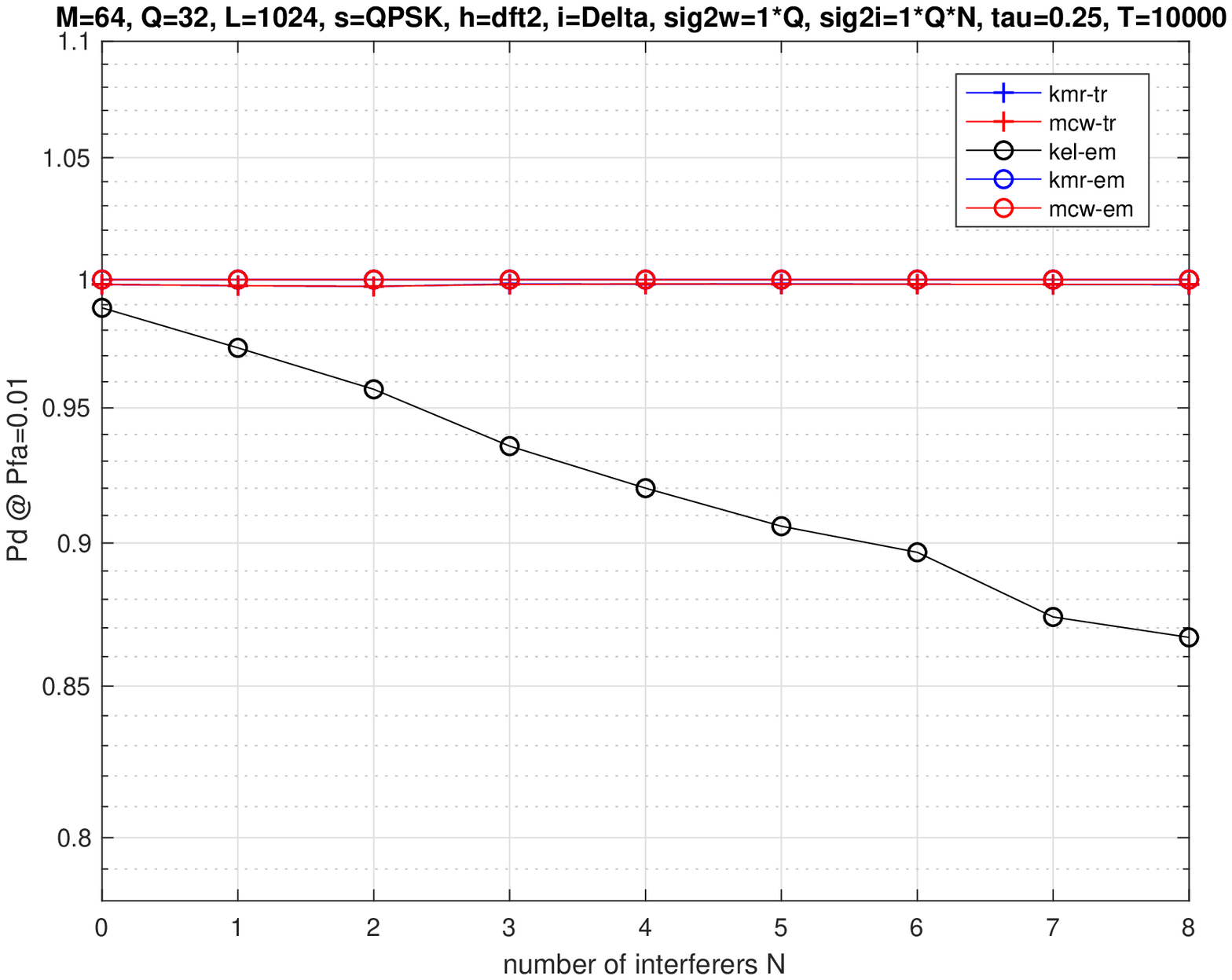}}{\putFrag{}}
        {\textb{\Metric versus number of interferers $N$ for various detectors, under
         $\nu=Q$, $\sigma_i^2=QN$,
         $M=64$, $Q=32$, $L=1024$, $N=5$, i.i.d.\ QPSK symbols, and
         spike-like interference.}}
        {\figsize}
        {\psfrag{Pr(err)}[B][B][0.8]{\sf $\Pr\{\text{error}\}$} 
         \psfrag{Pd @ Pfa=0.01}[B][B][0.8]{\sf \metric} 
         \psfrag{number of interferers N}[t][t][0.8]{\sf number of interferers $N$} 
         \psfrag{M=64, Q=32, L=1024, s=QPSK, h=dft2, i=Delta, sig2w=1*Q, sig2i=1*Q*N, tau=0.25, T=10000}{}}

\color{black}


\section{Conclusions}

In this paper, we considered the problem of detecting the presence/absence of a structured (i.e., partially known) signal from the space-time outputs of an array.
This problem arises when detecting communication signals, where often a few training symbols are known but the data portion is unknown apart from the symbol alphabet.
In our work, the signal's array response, the interference covariance, and the (white) noise variance are all assumed to be unknown.

We first reviewed GLRT-based detection of a known signal, highlighting previous work by Kelly \cite{Kelly:TAES:86} for full-rank interference, and by Kang/Monga/Rangaswamy \cite{Kang:TAES:14} and McWhorter \cite{McWhorter:ASAP:04} for low-rank interference with known rank $N$.
Next, we proposed EM-based extensions of these three detectors that apply to probabilistically structured signals, and
we established that the EM-based extension of Kelly's detector can be interpreted as ``soft'' version of Forsythe's iterative scheme from \cite[p.110]{Forsythe:LLJ:97}.
Finally, we proposed methods to estimate the interference rank $N$ when unknown, and we demonstrated the performance of our methods through numerical simulation.
The simulations showed that the error-rate of the proposed EM-based low-rank schemes was significantly lower than that of the training-based and/or full-rank schemes.

As future work, it would be interesting to consider the detection of multiple signals, as in \cite{Bliss:TSP:10}.
\textb{It would also be good to have a better theoretical understanding of how to do rank estimation and threshold selection for the proposed detectors.}


\section{Acknowledgments}

The authors thank Dr. Adam R.\ Margetts of MIT Lincoln Labs for inspiration, support, and insightful discussions.


\bibliographystyle{ieeetr}
\bibliography{macros_abbrev,books,misc,comm,radar,multicarrier,sparse,machine}


\def\baselinestretch{1.0}



\end{document}